\documentclass[acmtog]{acmart}
\acmSubmissionID{000}

\renewcommand\footnotetextcopyrightpermission[1]{} 

\usepackage[utf8]{inputenc}
\usepackage{booktabs} 
\usepackage[ruled]{algorithm2e} 
\usepackage{tabularx}
\usepackage{multirow}
\usepackage{subcaption}
\usepackage{textcomp}
\usepackage{color}
\usepackage{colortbl}
\usepackage{mathtools}
\usepackage{siunitx}
\usepackage{makecell}
\usepackage{bm}
\usepackage[usestackEOL]{stackengine}
\usepackage{stfloats}

\usepackage{xcolor} 
\usepackage{tikz}
\usetikzlibrary{fit,calc}

\makeatletter
\tikzset{%
  remember picture with id/.style={%
    remember picture,
    overlay,
    save picture id=#1,
  },
  save picture id/.code={%
    \edef\pgf@temp{#1}%
    \immediate\write\pgfutil@auxout{%
      \noexpand\savepointas{\pgf@temp}{\pgfpictureid}}%
  },
  if picture id/.code args={#1#2#3}{%
    \@ifundefined{save@pt@#1}{%
      \pgfkeysalso{#3}%
    }{
      \pgfkeysalso{#2}%
    }
  }
}

\def\savepointas#1#2{%
  \expandafter\gdef\csname save@pt@#1\endcsname{#2}%
}

\def\tmk@labeldef#1,#2\@nil{%
  \def\tmk@label{#1}%
  \def\tmk@def{#2}%
}

\tikzdeclarecoordinatesystem{pic}{%
  \pgfutil@in@,{#1}%
  \ifpgfutil@in@%
    \tmk@labeldef#1\@nil
  \else
    \tmk@labeldef#1,(0pt,0pt)\@nil
  \fi
  \@ifundefined{save@pt@\tmk@label}{%
    \tikz@scan@one@point\pgfutil@firstofone\tmk@def
  }{%
  \pgfsys@getposition{\csname save@pt@\tmk@label\endcsname}\save@orig@pic%
  \pgfsys@getposition{\pgfpictureid}\save@this@pic%
  \pgf@process{\pgfpointorigin\save@this@pic}%
  \pgf@xa=\pgf@x
  \pgf@ya=\pgf@y
  \pgf@process{\pgfpointorigin\save@orig@pic}%
  \advance\pgf@x by -\pgf@xa
  \advance\pgf@y by -\pgf@ya
  }%
}

\newcommand\tikzmark[2][]{%
\tikz[remember picture with id=#2] #1;}
\makeatother

\newcommand\MyBox[4][1.8ex]{%
  \tikz[remember picture,overlay,pin distance=0cm]
  {\draw[draw=#4,line width=1pt,fill=#4!20,rectangle,rounded corners, xshift=-7pt]
( $ (pic cs:#2) + (47ex,2ex) $ ) rectangle ( $ (pic cs:#3) + (1ex,#1) $ );
}
}

\usepackage{xcolor}
\usepackage{textcomp}
\usepackage{xspace}

\definecolor{gray}{rgb}{0.5,0.5,0.5}
\definecolor{purple}{rgb}{0.7,0.3,0.7}
\definecolor{blue}{rgb}{0,0,1}
\definecolor{darkblue}{rgb}{0,0,0.6}
\definecolor{orange}{rgb}{1,.5,0} 
\definecolor{red}{rgb}{1,0,0} 

\DeclarePairedDelimiterX{\norm}[1]{\lVert}{\rVert}{#1}
\DeclarePairedDelimiterX{\abs}[1]{\lvert}{\rvert}{#1}

\definecolor{MyDarkBlue}{rgb}{0,0.08,1}
\definecolor{MyDarkGreen}{rgb}{0.02,0.6,0.02}
\definecolor{MyDarkRed}{rgb}{0.8,0.02,0.02}
\definecolor{MyDarkOrange}{rgb}{0.70,0.35,0.02}
\definecolor{MyPurple}{RGB}{0.43,0,1.}
\definecolor{MyRed}{rgb}{1.0,0.0,0.0}
\definecolor{MyGold}{rgb}{0.75,0.6,0.12}
\definecolor{MyDarkgray}{rgb}{0.66, 0.66, 0.66}

\setcitestyle{square,nosort}

\SetAlFnt{\small}
\SetAlCapFnt{\small}
\SetAlCapNameFnt{\small}
\SetAlCapHSkip{0pt}
\IncMargin{-\parindent}

\setcopyright{none}
\acmJournal{TOG}
\acmYear{2021}\acmVolume{0}\acmNumber{0}\acmArticle{0}\acmMonth{1}

\newcommand{\ignore}[1]{}
\newcommand{\Comment}[1]{}

\newcommand{\ZX}[1]{\textcolor{blue}{[\textbf{Zexiang:} {\em #1}]}}

\newcommand{\MM}[1]{\textcolor{purple}{[\textbf{Mark:} {\em #1}]}}

\let\oldnl\nl
\newcommand{\nonl}{\renewcommand{\nl}{\let\nl\oldnl}}

\begin{document}

\title{Photon-Driven Neural Path Guiding}


\author{Shilin Zhu}
\affiliation{%
  \institution{University of California San Diego}
  \country{USA}}
\email{shz338@eng.ucsd.edu}

\author{Zexiang Xu}
\affiliation{%
  \institution{Adobe Research}
  \country{USA}}
\email{zexu@adobe.com}

\author{Tiancheng Sun}
\affiliation{%
  \institution{University of California San Diego}
  \country{USA}}
\email{tis037@eng.ucsd.edu}

\author{Alexandr Kuznetsov}
\affiliation{%
  \institution{University of California San Diego}
  \country{USA}}
\email{a1kuznet@eng.ucsd.edu}

\author{Mark Meyer}
\affiliation{%
  \institution{Pixar Animation Studios}
  \country{USA}}
\email{mmeyer@pixar.com}

\author{Henrik Wann Jensen}
\affiliation{%
  \institution{University of California San Diego and Luxion}
  \country{USA}}
\email{henrik@cs.ucsd.edu}

\author{Hao Su}
\affiliation{%
  \institution{University of California San Diego}
  \country{USA}}
\email{haosu@eng.ucsd.edu}

\author{Ravi Ramamoorthi}
\affiliation{%
  \institution{University of California San Diego}
  \country{USA}}
\email{ravir@cs.ucsd.edu}

\renewcommand{\shortauthors}{Zhu et al.}

\begin{abstract}
Although Monte Carlo path tracing is a simple and effective algorithm to synthesize photo-realistic images, it is often very slow to converge to noise-free results when involving complex global illumination. 
One of the most successful variance-reduction techniques is path guiding, which can learn better distributions for importance sampling to reduce pixel noise. 
However, previous methods require a large number of path samples to achieve reliable path guiding. 
We present a novel neural path guiding approach that can reconstruct high-quality sampling distributions for path guiding from a sparse set of samples, using an offline trained neural network.
We leverage photons traced from light sources as the input for sampling density reconstruction, which is highly effective for challenging scenes with strong global illumination.
To fully make use of our deep neural network,
we partition the scene space into an adaptive hierarchical grid, in which we apply our network to reconstruct high-quality sampling distributions for any local region in the scene.
This allows for highly efficient path guiding for any path bounce at any location in path tracing.
We demonstrate that our photon-driven neural path guiding method can generalize well on diverse challenging testing scenes that are not seen in training.
Our approach achieves significantly better rendering results of testing scenes than previous state-of-the-art path guiding methods.

\end{abstract}

\Comment{
However, previous path guiding methods require a large number of path samples to create high-enough desirable densities stored in different parts of the scene. In this paper, we develop the first path guiding system with a robust neural network that quickly reconstructs high-quality sampling densities for arbitrary bounces from sparse photons. By learning the shape of the target density function from the offline training data, our pre-trained neural network can be applied to new scenes. In addition, by caching the learned densities into an adaptive spatial grid based on hashing, we can guide path tracing efficiently. We implemented our system on Mitsuba CPU renderer with PyTorch C++ API that supports fast neural network inference on GPUs. Evaluations on multiple diverse test scenes have demonstrated better performance compared with state-of-the-art baseline path guiders. We believe our work sheds light on incorporating powerful deep learning techniques into the sampling density reconstruction in path guiding.}

%
%

%
%

\newcommand{\teaserwidth}{1.65in}
\begin{teaserfigure}
    \centering
  \includegraphics[width=\linewidth]{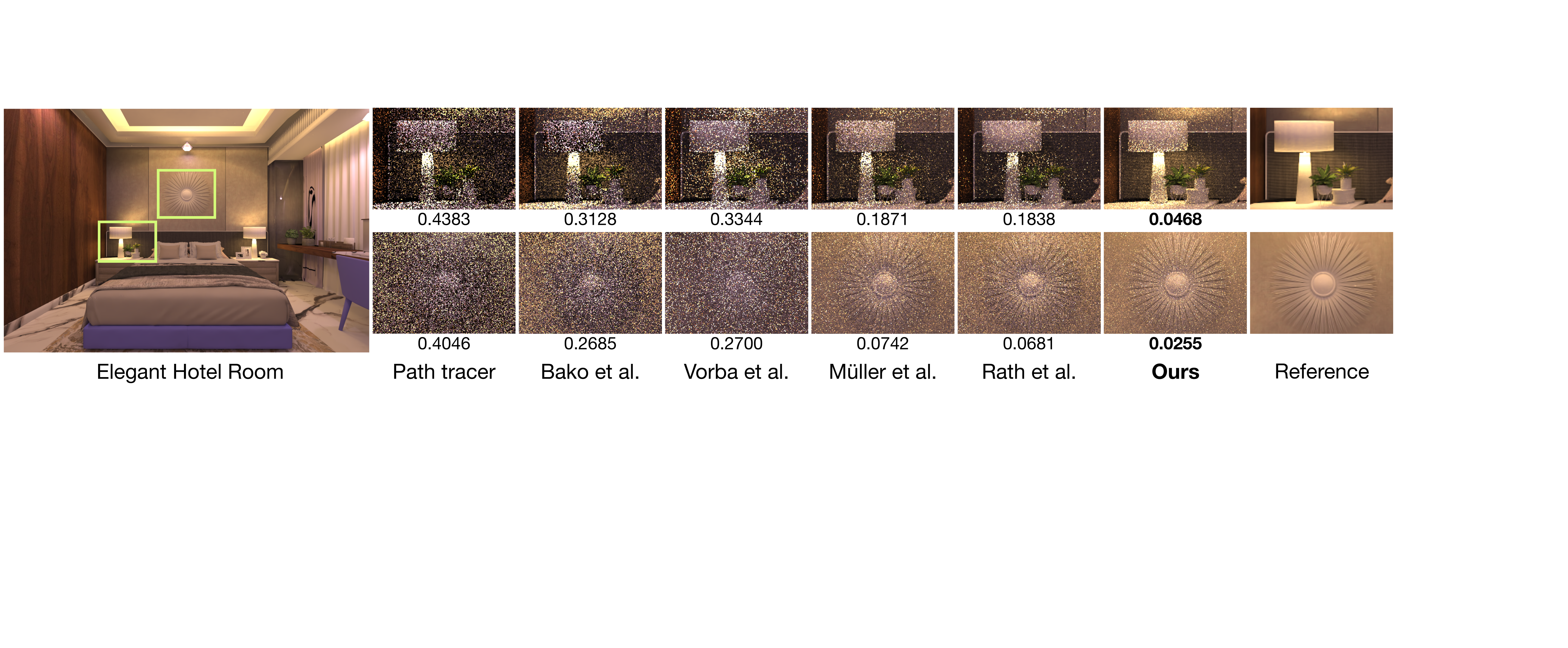}
  \caption{We present a novel photon-driven neural path guiding approach that can effectively reduce the variance of path tracing results. This complex scene is lit by several decorative ceiling lights which are extremely difficult to discover in path tracing. We compare the equal-time ($\sim$20 minutes) rendering results with standard path tracing and state-of-the-art path guiding methods (including \citet{muller2017practical}, \citet{bako2019offline}, and \citet{Rath2020}), showing the crops of the rendered results with corresponding relative MSEs (rMSEs).
  Recently, \citet{bako2019offline} use an offline trained neural network for path guiding; however, it only supports guiding the first bounce, which is not effective since this scene is dominated by indirect lighting. 
  On the other hand, while traditional methods allow for multi-bounce path guiding, they are purely online learning methods and it is highly expensive for them to learn the complex sampling functions for this challenging scene. 
  Our method utilizes an offline trained deep neural network and enables neural path guiding at any path bounces. Ours achieves the best rendering results qualitatively and quantitatively.}
    \label{fig:teaser}
\end{teaserfigure}

\maketitle

\section{Introduction}
Monte Carlo path tracing has been widely used in photo-realistic image synthesis.
However, while simple and flexible, path tracing can take a significant amount of time to generate noise-free images for complex scenes (e.g., Fig.~\ref{fig:teaser}).
One critical challenge for Monte Carlo based methods is to effectively construct light transport paths connecting the light and the camera.

Many path guiding methods \cite{muller2017practical,jensen1995importance} have been presented to construct advanced distributions (usually approximating incident light fields or some variants of those) for importance sampling at local shading points, 
guiding the local path sampling for high-energy path construction.
The recent successful ones are unidirectional guiding methods \cite{muller2017practical,Rath2020}; they rely on early path samples to discover high-energy sampling directions. 
However, this unidirectional path discovery process can be slow for a challenging scene that is dominated by indirect illumination.
While using light paths is known to be efficient in exploring the path space, previous photon-driven or bidirectional path guiding methods \cite{jensen1995importance,vorba2014line} are not yet efficient, requiring sampling a large number of light paths. 

We present a novel path guiding approach that can achieve highly efficient path sampling using only \emph{a sparse set} of light paths as input, thus significantly advancing the overall rendering speed.
Inspired by the original path guiding work \cite{jensen1995importance}, we leverage photons to compute local sampling distributions for importance sampling in path tracing.
As is done by \citet{jensen1995importance}, a sampling distribution at any 3D local region can be easily obtained by binning local photons according to their directions (i.e., a 2D histogram map).
However, such distributions are only reliable with locally dense enough photons, and, on the other hand, are usually low-quality and appear highly noisy with sparse photons (see Figs.~\ref{fig:net} and~\ref{fig:map_vis}). 

We propose to use a deep neural network to reconstruct high-quality sampling maps for path guiding from low-quality noisy sampling maps that are acquired by binning sparse photons (see Fig.~\ref{fig:net}). Our approach is \emph{the first deep learning based photon-driven path guiding approach}. In essence, we break down the complex path guiding problem, mainly focusing on reconstructing local sampling distributions represented as 2D maps (i.e., images), and thus pose this problem as one of the image-to-image reconstructions that can now be addressed by deep learning techniques.
Our sampling map reconstruction network is effectively trained offline in a scene-independent way. 
The trained network is able to recover the accurate shapes of a diverse set of complex sampling distributions on challenging novel scenes, which enables highly efficient guided path tracing with complex global illumination effects.

Our network is designed to reconstruct high-quality sampling maps at local spatial regions. 
To make these sampling maps well distributed and locally representative in the scene space, 
we adaptively partition the entire scene space into a hierarchical grid, according to the complexity of local geometry and incident light.
The sampling map of every leaf voxel in the grid is reconstructed by our network, enabling path guiding at any location in a scene. 
Note that, our approach allows for efficient guided path sampling at any bounce points; this is \emph{the first offline-learning neural path guiding approach that can guide arbitrary bounces}.
We demonstrate that our novel deep path guiding achieves significantly better rendering quality on various challenging scenes than previous state-of-the-art path guiding methods given equal rendering time (see Fig.~\ref{fig:teaser}).

In summary, our main contributions are:
\begin{itemize}

\item We present the first deep learning based photon-driven path guiding approach;
\item To our knowledge, this is the first offline-learning neural path guiding approach that can guide arbitrary bounces;
\item Our proposed framework generalizes well to unseen new scenes and produces significantly better rendering results.

\end{itemize}

\section{Related Work}
\label{sec:related}


\paragraph{Monte Carlo rendering.}
One central problem of computer graphics is to efficiently evaluate the rendering equation~\cite{kajiya1986rendering}, which describes how light transports globally inside a scene. Monte Carlo methods are among the most effective methods to compute the light transport, which require effectively sampling high-energy paths that connect the camera and light for efficient rendering.
Since Monte Carlo path tracing was introduced in the seminal work by~\citet{kajiya1986rendering}, numerous papers have developed more efficient methods to explore path space, including bidirectional path tracing~\cite{lafortune1993bi,veach1995bidirectional} and metropolis light transport~\cite{veach1997metropolis,pauly2000metropolis}. These methods typically leverage importance sampling to sample sub-path directions at any bounces for each traced path traversing the scene.
Since the incident illumination is unknown, the importance sampling usually only considers the reflectance term (with a cosine term) in the rendering equation (please refer to Sec.~\ref{sec:background} for more details);
this however is not efficient for challenging scenes with complex indirect lighting.
Path guiding \cite{jensen1995importance,vorba19guiding} can instead provide more efficient importance sampling; our novel photon-driven path guiding approach can reconstruct high-quality sampling distributions that well approximate the complex incident light fields, thus leading to highly efficient rendering.

\paragraph{Photon-based rendering.} 
Particle density estimation has also been applied in computer graphics to evaluate the rendering equation, which introduces photon mapping and many other particle- or photon- based rendering methods \cite{shirley1995global,jensen1996global,hachisuka2008progressive,knaus2011progressive}.
These methods focus on photon density estimation at any given shading point, which avoids the high-frequency noise in MC rendering and is very effective for computing complex global illumination.
Photon density estimation can only provide biased radiance or irradiance estimates, since it blurs the photon contributions within a certain kernel bandwidth (though this bias can be consistently reduced to zero by progressively reducing the bandwidth and tracing infinite photons \cite{hachisuka2008progressive,hachisuka2009stochastic,knaus2011progressive}).
Our goal is not to compute photon density for a single point but to approximate incident light fields for a local area (in a voxel) as sampling distributions.
Therefore, we consider the integral of irradiance over an area (i.e., the incident flux), which can be effectively evaluated using photons in an unbiased way. 

Recently, \citet{zhu2020deep} introduce a deep learning based method for photon density estimation in photon mapping. \Comment{\MM{Have you compared against Zhu?}} They leverage a PointNet \cite{qi2017pointnet} style neural network to process individual photons. However, the complexity of running such a network grows linearly with the number of photons. 
We instead leverage a UNet \cite{ronneberger2015u} style network and consider a raw photon histogram map, composed by binning photons \cite{jensen1995importance}, as input; 
therefore, the complexity of our network is independent to the photon count and runs in constant time.
We show that our method consistently reconstructs better sampling distributions with more photons. 

\paragraph{Path guiding.}
In general, path guiding aims to estimate the incoming light fields and draw samples accordingly to accelerate the convergence of Monte-Carlo rendering. 
The first path guiding technique is based on photons \cite{jensen1995importance}; it traces light paths from the light sources, distributes photons in the scene, and constructs local photon histograms as sampling distributions for the importance sampling in path tracing. 
Though very efficient to compute, such histogram-based sampling maps are only of high quality when accumulating dense enough photons. 
We extend this simple classical histogram-based technique to a novel learning-based method in a new path guiding framework;
our method regresses high-quality sampling maps from sparse photons, avoiding expensively tracing a large number of photons.

\citet{vorba2014line} present a bidirectional guiding method, where both camera paths and light paths are guided using online fitted gaussian-mixture (GM) distributions at spatial cache points. 
This technique was further extended to product sampling~\cite{herholz2016product}, and to account for parallax~\cite{ruppert2020robust}. 
However, the online fitting process in these methods is usually slow and the GM model also makes it difficult to express high-frequency sampling distributions. 
Our approach leverages histograms as input (that can be easily computed at very low cost online) and an offline trained compact neural network that can rapidly reconstruct high-quality sampling maps with high-frequency details from the input.

Recently, unidirectional guiding methods have become more effective and practical, thanks to the efficient adaptive guiding framework introduced by \citet{muller2017practical}.
Many works extend this framework to achieve sampling in primary space \cite{guo2018primary}, product sampling \cite{diolatzis2020practical}, and variance-aware sampling \cite{Rath2020}.
These methods iteratively trace camera paths to adaptively reconstruct the incident light fields; 
this relies on early iteration paths to discover the light sources, in order to reconstruct reliable sampling distributions to guide the following iteration paths. 
However, the light discovery can be slow and unsuccessful for a scene with dominant indirect lighting, and the errors in the early-iteration sampling distributions can bias the path sampling in later iterations and never get fixed.
In contrast, we leverage photons that are efficient in exploring indirect light transport; our learning based approach can also recover high-quality sampling distributions from sparse photons at an early stage, effectively avoiding a slow start in the guiding and rendering.
Moreover, our photons are traced independently in each iteration, which avoids accumulating the sampling errors through multiple iterations.

\paragraph{Neural path guiding.}
Recently, deep learning techniques have been applied in path guiding.
\citet{muller2019neural} train an online neural network to perform importance sampling in global path space. 
This method can reproduce accurate ground-truth sampling functions, but the online training process is extremely slow. 
Some recent works leverage offline trained networks \cite{bako2019offline,huo2020adaptive}; however, they only guide the path sampling at the first bounce.
While we also leverage an offline trained neural network, our method instead leverages photons and supports guiding at any bounces, enabling significantly better rendering results than the first-bounce guiding approach \cite{bako2019offline} (see Fig.~\ref{fig:teaser}).

\newcommand{\Px}{\bm{x}}
\newcommand{\DirO}{\omega_{o}}
\newcommand{\DirI}{\omega_{i}}
\newcommand{\DeltaO}{\Delta\Omega}
\newcommand{\DeltaA}{{\Delta A}}
\newcommand{\PathN}{{N_c}}
\newcommand{\PathF}{{N_f}}
\newcommand{\PhotonN}{{N_p}}
\newcommand{\PhotonVoxelM}{{M}}
\newcommand{\PathVoxelM}{{Q}}
\newcommand{\PhotonNormalVar}{{V_n}}
\newcommand{\SMap}{S}

\section{Background}
\label{sec:background}


Physically-based rendering can be expressed by the Rendering Equation \cite{kajiya1986rendering} that describes the radiance leaving an intersection point $\Px$ in direction $\DirO$:
\begin{equation}
    L(\Px, \DirO) = L_{e}(\Px, \DirO) + \int_{\Omega}L_{i}(\Px, \DirI)f_{r}(\Px, \DirI, \DirO)\cos\theta_{i}d\DirI,
    \label{eqn:re}
\end{equation}
where $L_{e}(\Px, \DirO)$ denotes the emitted radiance, $L_{i}(\Px, \DirI)$ is the incident radiance from direction $\DirI$, $f_{r}$ is the bidirectional scattering distribution function (BSDF), and $\Omega$ is the visible hemisphere. The key component in the equation is the integral that computes the reflected radiance $L_r(\Px, \DirO) = \int_{\Omega}L_{i}(\Px, \DirI)f_{r}(\Px, \DirI, \DirO)\cos\theta_{i}d\DirI$ over all directions in the hemisphere. 

The integral can be numerically evaluated using Monte Carlo estimation \cite{veach1997robust}:
\begin{equation}
    L_{r}(\Px, \DirO) = \frac{1}{N} \sum_{i=1}^{N} \frac{L_{i}(\Px, \DirI)f_{r}(\Px, \DirI, \DirO)\cos\theta_{i}}{p(\DirI)}
    \label{eqn:mc}
\end{equation}
where $N$ Monte Carlo path samples in various directions $\DirI$ are drawn from the probability density function (PDF) $p(\DirI)$. 
Considering global illumination with multiple bounces, $L_{i}(\Px, \DirI)$ is in fact computed by recursively evaluating integrals using Eqn.~\ref{eqn:re}.
Therefore in Monte Carlo path tracing, rays are sampled from each intersection point to compute the radiance that contributes to the pixel color at multiple bounces. 

The variance of the Monte Carlo estimate $L_{r}(\Px, \DirO)$ can be reduced by sampling $\DirI$ from a density function $p(\DirI)$ that resembles the numerator $L_{i}(\Px, \DirI)f_{r}(\Px, \DirI, \DirO)\cos\theta_{i}$. Ideally, if $p(\DirI)$ and the numerator only differ by a constant scale, the variance is reduced to zero. However, this numerator is unknown and is as difficult as the integral to compute, due to complex visibility and indirect lighting in $L_i$; therefore, in practice, path tracing often proceeds with BSDF importance sampling. 

Path guiding aims to reconstruct a density function that matches the shape of the numerator as closely as possible. In particular, since the standard BSDF importance sampling satisfies 
$$p_{\text{BSDF}}(\DirI) \propto f_{r}(\Px, \DirI, \DirO),$$ 
recent path guiding methods often set the target probability density to be proportional to the incident light \cite{vorba2014line,muller2017practical}
\begin{equation}
    p_{\text{guide}}(\DirI) \propto L_{i}(\Px,   \DirI)\cos\theta_{i}.
    \label{eqn:target}
\end{equation}
The final sampling strategy is achieved by combining the guiding and BSDF sampling using one-sample Multiple Importance Sampling (MIS): \cite{veach1995optimally}
\begin{equation}
    p(\DirI) = \alpha p_{\text{BSDF}}(\DirI) + (1-\alpha)p_{\text{guide}}(\DirI),
    \label{eqn:target_mis}
\end{equation}
where $\alpha$ is the mixture coefficient that determines the probability of choosing BSDF sampling or guided sampling.

Many recent works rely on early path samples in the path tracing to approximate the incident light field (Eqn.~\ref{eqn:target}), which is not sufficient for challenging scenes with strong indirect illumination as shown in Fig.~\ref{fig:teaser}.
We instead leverage photons traced from the light sources to compute the sampling density functions, which effectively explores the challenging light transport.
Our novel approach advances the traditional path guiding with powerful deep learning techniques and an efficient spatial structure, thus enabling highly efficient path guiding from sparse photons.

\begin{figure*}[t]
    \includegraphics[width=\linewidth]{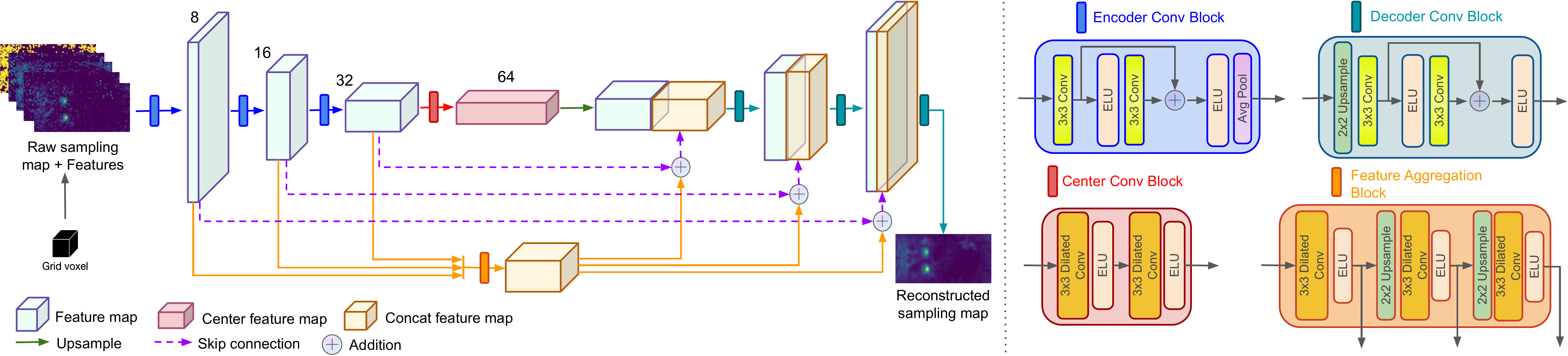}
    \vspace{-5mm}
    \caption{The neural network architecture for sampling map reconstruction. We use a compact autoencoder with light-weight masked convolutions \cite{liu2018image, yi2020contextual} and ELU \cite{clevert2015fast} activation function which can extract high-level features from the input energy map and output a smooth and dense sampling map. The bottleneck layers use dilated convolutions \cite{iizuka2017globally} to further expand the size of the receptive fields. \Comment{The text and sampling maps in this figure is too small... Maybe reformat or even split it to two figures? Also maybe improve the illustration of the skip links.}}
    \label{fig:net}
    \vspace{-3mm}
\end{figure*}

\section{Overview}
\label{sec:overivew}
Our path guiding approach uses a deep neural network to regress high-quality sampling maps that can be used to guide path sampling. Correspondingly, we introduce a novel practical path guiding framework that utilizes our neural network to reconstruct sampling maps in an adaptive spatial hierarchical grid, enabling effective path guiding at multiple bounces.
In the following sections, we first introduce our sampling map parameterization, target sampling density, and how to use photons to compute the sampling maps in Sec.~\ref{sec:photoncomputing}.
We then introduce our deep neural network that can regress high-quality sampling maps given noisy low-quality sampling maps in Sec.~\ref{sec:learning}.
We present our full neural path guiding framework in Sec.~\ref{sec:neuralguiding}, which describes our iterative guiding and rendering process, adaptive spatial structure, and how paths, photons, and the neural network are incorporated in the system.
The implementation details are discussed in Sec.~\ref{sec:impl}.
We present an extensive evaluation of our method in Sec.~\ref{sec:eval}.
In the end, we conclude our paper and discuss future work in Sec.~\ref{sec:future}.

\section{Computing sampling maps from photons}
\label{sec:photoncomputing}

Previous methods \cite{jensen1995importance,vorba2014line} usually compute hemispherical distributions at sampled surface points to approximate incident light fields. However, such hemispherical functions only approximate light fields at very local flat 2D surface regions, and are hard to interpolate on surfaces with complex normal variation. 
Inspired by the recent unidirectional path guiding methods \cite{muller2017practical,Rath2020,bako2019offline},
we utilize a full spherical sampling distribution (instead of a hemispherical one) that models the incident light distribution in a local 3D region.
In particular, we build a hierarchical grid (see Sec.~\ref{sec:grid}) in the scene space, and compute a spherical sampling distribution for each local 3D voxel of the grid.  
In this section, we discuss the representation of our sampling function and the computation of it from photons.

\paragraph{Spherical function representation.}
We use a regular directional grid that represents the sampling density function as a 2D sampling map (similar to \cite{bako2019offline}). 
We leverage the cylindrical mapping to parameterize the spherical domain for better area preservation (similar to \cite{muller2017practical}). In particular, a unit vector $r=(x, y, z)$ (corresponding to a point on a unit sphere) is mapped to a 2D location $ (u, v) = (z, \phi)$ on the sampling map, where $\phi=\arctan(y/x)$.
Our sampling map is like a standard environment map or radiance map in lighting representation, but ours is monochromatic and uses cylindrical mapping.

\paragraph{Target sampling density.}
As discussed in Sec.~\ref{sec:background} (Eqn.~\ref{eqn:target}), in general, the goal of path-guiding is to compute the sampling density at any position, making it proportional to the incident light $L_{i}(\Px, \DirI)\cos\theta_{i}$. 
For our discrete case where we consider a 3D voxel region and a certain pixel range (representing a solid angle bin) of a sampling map, 
it is in fact the expected incident light that is of our interest.
In particular, given a voxel $j$ and a solid angle footprint of pixel $k$ in the sampling map, the expected $L_{i}(\Px, \DirI)\cos\theta_{i}$ coming from the solid angle over the 2D surface area (that is of the scene geometry located in the voxel) inside the voxel is expressed by:
\begin{align}
    \mathbb{E}(L_{i}(\Px, \DirI)\cos\theta_{i})  
    & =\frac{\int_{\DeltaA_j}\int_{\DeltaO_k} L_{i}(\Px, \DirI) \cos\theta_{i}d\DirI d\Px} {\DeltaO_p \DeltaA_j} \\
    & =\frac{\Phi_{j,k}} {\DeltaO_p \DeltaA_j},
    \label{eqn:expected_sampling}
\end{align}
where $\DeltaA_j$ represents the entire surface area of the scene geometry covered by the voxel $j$, $\DeltaO_k$ represents the solid angle footprint covered by the pixel $k$ in the sampling map, and $\Phi_{j,k}$ represents the total incident energy in the spatial and directional range.
Therefore, it is the total energy (radiant flux)
\begin{align}
    \Phi_{j,k} 
    & =\int_{\DeltaA_j}\int_{\DeltaO_k} L_{i}(\Px, \DirI) \cos\theta_{i}d\DirI d\Px,
    \label{eqn:samplingenergy}
\end{align}
that governs our sampling map distribution.
Essentially, $\Phi_{j,k}$ models the integrated incident radiance and is proportional to the sampling probability of a pixel $k$ in a sampling map at a voxel $j$. 
Note that, the irradiance ($E(\Px, \DeltaO_k) = \int_{\DeltaO_k} L_{i}(\Px, \DirI) \cos\theta_{i}d\DirI$) at surface point $x$ is a standard radiometry term and widely discussed in previous works \cite{jensen1995importance,Rath2020}; when divided by the total area, $\Phi_{j,k}$ also describes the expected irradiance ($\Phi_{j,k}/\DeltaA_j$) in the voxel. 
Therefore, we seek to obtain sampling densities that are proportional to the expected incident light:
\begin{equation}
    p_{\text{guide}}(\DirI) \propto \Phi_{j,k_i}/\DeltaO_{k_i},
    \label{eqn:energy_target}
\end{equation}
where $k_i$ is the pixel covering direction $\DirI$ in the sampling map, and we ignore the $\DeltaA_j$ in Eqn.~\ref{eqn:expected_sampling} since it is a constant value for all solid angles in a voxel.
This sampling density corresponds to a sampling map, each pixel value of which is proportional to $\Phi_{j,k_i}$.
We thus reconstruct a sampling map by normalizing an energy map that records the energy $\Phi_{j,k_i}$ in each pixel.


\paragraph{Computing incident energy.}
In this work, we leverage particle tracing to effectively evaluate the integral of $\Phi_{j,k}$ (Eqn.~\ref{eqn:samplingenergy}). 
\Comment{
Similar to \cite{muller2017practical}, path tracing is used to accumulate irradiance samples $E_{j,k}$ inside a 3D region to approximate Eqn.~\ref{eqn:samplingenergy}:
\begin{equation}
    \Phi_{j,k} \propto E_{j,k} = \frac{1}{N_{\DeltaO_k}}\sum_{\omega_p \in \DeltaO_k, \Px_p \in \DeltaA_j}\frac{L_{i}(\Px, \DirI) \cos\theta_{i}}{p(\omega_{i})}
    \label{eqn:path}
\end{equation}
}
We trace light paths from the light sources to distribute photons in the scene, where each photon carries a portion of flux; 
$\Phi_{j,k}$ can then be evaluated by simply binning the photons similar to \cite{jensen1995importance}.
In particular, $\Phi_{j,k}$ is estimated by:
\begin{equation}
    \Phi_{j,k} = \sum_{\omega_p \in \DeltaO_k, \Px_p \in \DeltaA_j} \Delta \Phi_{p},
    \label{eqn:photon}
\end{equation}
where $p$ denotes a photon, the photon arrives at the surface point $\Px_p$ from direction $\omega_p$, and $\Delta \Phi_{p}$ is the energy carried by the photon. 
Equation~\ref{eqn:photon} essentially accumulates all the photon energies inside the voxel and directional bin.

Note that, \citet{muller2017practical} leverages path tracing to accumulate the radiance samples inside a 3D region; this can also be seen as an integral (a Monte Carlo one) of the radiance over an area and a solid angle, similar to the energy integral of Eqn.~\ref{eqn:samplingenergy}. 
We leverage photon tracing to evaluate the integral and our particle-based approach provides an unbiased estimate for the energy $\Phi_{j,k}$ when the photon count goes to infinity.

Since the evaluation is governed by a single summation, we can progressively trace as many photons as needed, and accumulate the photons to compute an energy map without any memory bottleneck. 
Once a photon is accumulated in a voxel, the photon data is immediately deleted, except when the grid needs to be refined at the beginning (Sec.~\ref{sec:grid}).
Note that, an accurate energy map requires tracing a large number of photons, but in practice, we can only allow for tracing a small number of photons at rendering time, which by themselves cannot directly lead to high-quality sampling. 
We propose to compute high-quality sampling maps, using a large number of photons, and take them as ground truth to train a deep neural network offline that can regress high-quality sampling maps online efficiently.

\begin{figure*}[t]
    \includegraphics[width=\linewidth]{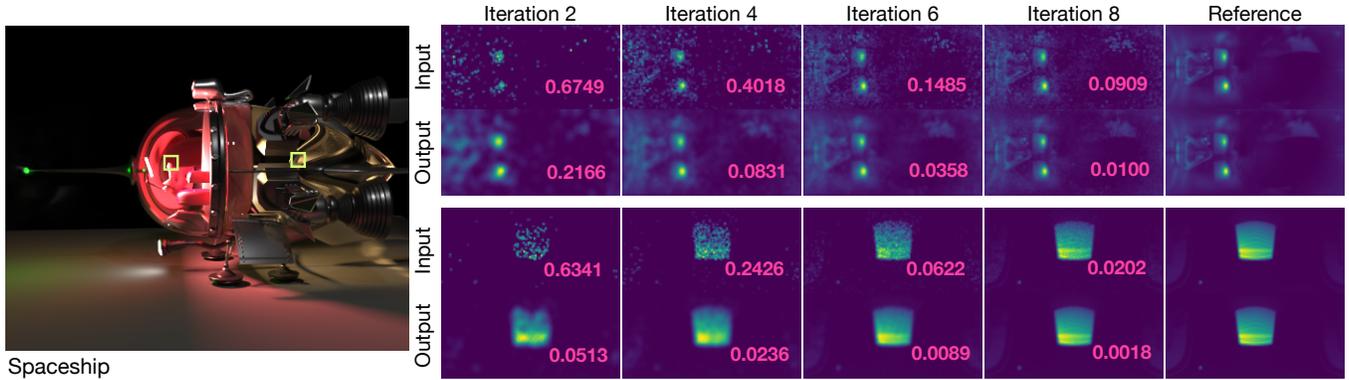}
    \caption{Example input and output sampling maps of the pre-trained neural network over iterations (gamma transformed for better visualization purpose). With more iterations of path and photon tracing, both the input raw sampling map and the reconstructed output sampling map get better over time. Numbers are rMSE computed using the reference sampling maps.
    \Comment{\MM{label should be input and output (not Raw/Rec) or you should define Raw/Rec int the caption.  May also want to describe what 'iteration' means since it is an overloaded term}}}
    \label{fig:map_vis}
\end{figure*}

\section{Learning to regress high-quality sampling maps}
\label{sec:learning}
While using a large number of photons can result in an accurate estimate of $\Phi_{j,k}$, it requires a significant amount of tracing time. 
On the other hand, computing a sampling map from sparse photons is fast, but the map is usually low-quality and appears noisy with many empty bins. 
As a result, neither using dense photons (too slow) nor sparse photons (too low-quality) is suitable for efficient path guiding. 
To overcome this, our central idea is to obtain accurate sampling maps offline as ground truth using dense enough photons, and then leverage supervised learning to regress such maps from low-quality maps that can be computed efficiently from sparse photons for path guiding. 
Specifically, we propose to train a deep convolutional neural network that learns to reconstruct a high-quality sampling map from sparse photons.

Our sampling maps are reconstructed iteratively through multiple iterations in our path guiding framework. 
Specifically, we consider a raw sampling map $\SMap_{e,t}$ (1 channel) as input, acquired by accumulating a sparse set of traced photons from iteration 1 to $t$ using Eqn.~\ref{eqn:photon}, where $t$ denotes the iteration number and $e$ means accumulated photon energy. 
To give the neural network a hint on how the raw sampling map evolves over previous iterations with more photons, we keep the raw sampling map $\SMap_{e,t-1}$ from the previous iteration also as an input channel.
In addition, we record the number of photons per solid angle bin in $\SMap_{e,t}$ and $\SMap_{e,t-1}$, resulting in two additional maps $P_{e,t}$ and $P_{e,t-1}$, and use them as auxiliary buffers in the input, which provides two additional input channels. 
Inspired by the image inpainting techniques \cite{liu2018image, yu2019free, yi2020contextual}, we also concatenate a binary mask $B_{e,t}$ (1 channel) indicating whether a solid angle bin contains photon data or not, and use light-weight masked convolutions to process the input maps. 
As a result, our full input is a 2D image map with 5 channels in total and our network $\mathbb{F}$ can be expressed by:
\begin{equation}
    \SMap_{d} = \mathbb{F}(\SMap_{e,t}, \SMap_{e,t-1}, P_{e,t}, P_{e,t-1}, B_{e,t} ).
\end{equation}
Our network learns to regress a one-channel sampling map $\SMap_{d}$, supervised by the ground-truth map $\tilde{\SMap_{d}}$ computed from a large number of photons.

\subsection{Network architecture} 
Note that, our network is essentially designed to solve an image-to-image reconstruction task. 
Many existing 2D neural networks for image-to-image denoising, translation, and impainting  (\cite{chaitanya2017interactive, bako2017kernel, vogels2018denoising,liu2018image}) can thus be potentially applied to address our problem.
However, our network is applied on a large number (thousands) of voxels, while our end goal is to speed up the total rendering process.
Therefore, we balance the inference speed and reconstruction quality in our network design.

We propose to use a compact U-Net \cite{ronneberger2015u} style neural network with residual links and skip connections to achieve the sampling map reconstruction as illustrated in Fig.~\ref{fig:net}. 
Our network uses multiple downsampling and upsampling 
convolutional layers to extract meaningful neural features from the input sampling map $\SMap_{e}$ and regress a high-quality sampling map $\SMap_{d}$. 
Our input raw sampling maps are computed from sparse photons, which contain many holes or empty bins as shown in Fig.~\ref{fig:map_vis}.
Therefore, we use the light-weight masked convolutions as the convolutional layers in our network, inspired by the recent image inpainting works \cite{liu2018image, yi2020contextual}.
This ensures that valid (non-empty) and invalid (empty) solid angle bins are treated differently in the network and only valid bins can contribute in a convolutional operation. 
Note that, our network is relatively compact, compared to the previous U-Net-like networks (\cite{chaitanya2017interactive, bako2017kernel, vogels2018denoising}) used in other tasks; the maximum number of feature channels in our network is only 64.
This compactness allows for fast sampling map reconstruction during path guiding on high-end GPUs, keeping our network from becoming the bottleneck in the entire rendering process. 
In fact, a large network is not very necessary for our task, since a sampling map has only a single channel (no color variations) and we only need to reconstruct low-resolution maps (just $32\times64$ or $64\times128$ that are much lower than other reconstruction tasks), which are already adequate for high-quality rendering. 
While compact, our network can regress high-quality maps that enable efficient path guiding in path tracing with quickly reduced variances. We believe our network size can be further reduced by advanced network compression techniques \cite{cheng2018model, deng2020model} that can enable even more efficient path guiding, and we leave this as future work.

We also find that the usage of photon counts and the previous sampling map as buffers is effective; these auxiliary features are simple to obtain but useful indicators for the quality and evolution of the original per-solid-angle probabilities. 
These buffers are also compact and improve the reconstruction quality with marginal extra cost. On the other hand, we find that other geometric features, such as position, normal, and depth -- that are used in previous screen-space guiding methods \cite{bako2019offline, huo2020adaptive} -- are not very helpful in most of our scenes since our reconstruction operates on each local 3D voxel. To justify our neural network design, we compare the performance with a standard U-Net \cite{ronneberger2015u} in the supplementary material.
\Comment{
\MM{It would be great to see an ablation of the network with and without the aux buffers (photon counts and previous buffers - both and each individually) - is that what the UNet test in fig 13 is supposed to be?  If so, maybe reference it here.  Showing a comparison with the position/normal/depth vs not would also be interesting.}}

\colorlet{pink2}{red!50}
\colorlet{blue2}{cyan!50}
\colorlet{green2}{green!50}
\colorlet{purple2}{purple!50}

\MyBox{starta}{enda}{green2}
\MyBox{startb}{endb}{pink2}
\MyBox{startc}{endc}{green2}
\MyBox{startd}{endd}{pink2}
\MyBox{starte}{ende}{blue2}
\MyBox{startf}{endf}{green2}
\MyBox{startg}{endg}{blue2}
\MyBox{starth}{endh}{purple2}

\LinesNumbered

\begin{algorithm}
    \SetAlgoLined
    \SetKwInOut{Input}{Input}
    \SetKwInOut{Output}{Output}
    \Input{Target scene, pre-trained neural network $\mathbb{F}$}
    \Output{A rendered image}
    \tikzmark{starta}
    Initialize a regular spatial grid; set all $\PathVoxelM_j=0$ \; 
    \tikzmark{enda}
    \For{ {\upshape each iteration} $t < T$} 
    {
    \textbf{Initiate $2^{t}$ SPP path samples}\; 
    \For{each path}{
    \For{{\upshape each bounce} $b$ }{
    Locate voxel $j$ ($\Px_b \in \DeltaA_j$) \;
    \tikzmark{startb}
    \eIf{{\upshape not isValid($j$)} (no sampling map)} 
        {
        $\text{Sample}(p_\text{BSDF})\rightarrow \omega_{b}$ \;
        }
        {
        $\text{Sample}(p_\text{MIS})\rightarrow \omega_{b}$  (Eqn.~\ref{eqn:mis_coef})\;
        }
        
        \tikzmark{endb}
        \tikzmark{startc}{markValid($j$) \;}
        \nonl\tikzmark{endc}\vskip-7pt
    }
    \textbf{Compute path throughput and $L(\Px_b, \omega_b)$ }\;
    \For{{\upshape each bounce} at $\Px_b\in \DeltaA_j$ }{
        \If{{\upshape isValid($j$)}}{
        \tikzmark{startd}
        $\nu_b = L(\Px_b, \omega_b) \cos \theta_b f_{r}(\Px, \omega_b, \DirO)$ \;
        \textbf{if} $\omega_b \leftarrow p_\text{guide}$ \textbf{then}
        $\nu_{j,\text{G}} \mathrel{+}= \nu_b$ \lElse{$\nu_{j,\text{B}} \mathrel{+}= \nu_b$ }\label{alg:nu}
        \textbf{if} $\omega_b \leftarrow p_\text{guide}$ \textbf{then}
        $\PathVoxelM_{j,\text{G}} \mathrel{+}= 1$ \lElse{$\PathVoxelM_{j,\text{B}} \mathrel{+}= 1$ }
       \Comment{$\PathVoxelM_j\mathrel{+}=1$ ;}\lIf{$\PathVoxelM_{j,\text{G}} \geq 50$ \text{\&} $\PathVoxelM_{j,\text{B}} \geq 50$}{Update $\alpha_j$ (Eqn.~\ref{eqn:mis_alpha})}
       \nonl\tikzmark{endd}\vskip-7pt
        }
        
    }
    Update the output image \;
    }

    \textbf{Trace $2^{t} \PhotonN$ light paths for photons}\;
    
    \For{each photon $p$}{
    Locate voxel $j$, solid angle $k$ ($\Px_p \in \DeltaA_j$, $\omega_p \in \DeltaO_k$) \;
        \If{{\upshape  isValid($j$)}}{
        \tikzmark{starte}
        Update energy map: $\Phi_{j, k} \mathrel{+}= \Delta\Phi_{p}$ (for Eqn.\ref{eqn:photon})\;
        \tikzmark{ende}
        \tikzmark{startf}
        $\PhotonVoxelM_j \mathrel{+}= 1$; Update $\PhotonNormalVar$ \;
        \If{$\PhotonVoxelM_j > \PhotonVoxelM_{\text{thr}}$ or $\PhotonNormalVar > \PhotonNormalVar_{\text{thr}}$}{
            Subdivide voxel $j$ into two sub-voxels (Sec.~\ref{sec:grid})\;
            }
            \nonl\tikzmark{endf}\vskip-7pt
        }
    }
    \For{each valid voxel $j$}{
        \tikzmark{startg}
        Reconstruct sampling maps (i.e. $p_{\text{Guiding}}$) with $\mathbb{F}$ \;
        \nonl\tikzmark{endg}\vskip-7pt}
    }
    \tikzmark{starth}
    \textbf{Trace $N_{f}$ paths for final output (Sec.~\ref{sec:finaltracing})}\;
    \nonl\tikzmark{endh}\vskip-7pt
    \caption{Our neural path guiding framework in Sec.~\ref{sec:neuralguiding}. 
    Through multiple iterations of path tracing and photon tracing, we construct a hierarchical grid (Sec.~\ref{sec:grid}), reconstruct and update the sampling map in each valid grid voxel (Sec.~\ref{sec:reconstruction}), and
    guide the path tracing using the sampling maps (Sec.~\ref{sec:guiding}).
    We also apply a final path tracing pass guided by the reconstructed sampling maps (Sec.~\ref{sec:finaltracing}).
    We use different colors to mark different subsections, with green for Sec.~\ref{sec:grid}, blue for Sec.~\ref{sec:reconstruction}, red for Sec.~\ref{sec:guiding} and purple for Sec.~\ref{sec:finaltracing}.  }
    \label{algo_pseudo}
\end{algorithm}

\subsection{Loss function}
We utilize an $L_{1}$ loss to supervise the output sampling map:
\begin{equation}
    \mathbb{L}_{\SMap} = |\hat{\SMap_{d}} - \SMap_{d}|
    \label{eqn:loss}
\end{equation}
where $\hat{\SMap_{d}}$ is the ground-truth sampling map computed by tracing a large number of photons. Inspired by the deep supervision in machine learning \cite{xie2015holistically, lee2015deeply}, we also provide the ground-truth signal to every decoding level in order to ease the loss backpropagation. 
To avoid potential over-blurring, we leverage an asymmetric function inspired by \cite{vogels2018denoising}; this leads to our full loss
\begin{equation}
    \mathbb{L}_{\text{rec}} = \mathbb{L}_{\SMap} \cdot (1 + (\lambda - 1) \cdot \mathbb{H})
    \label{eqn:loss_asym}
\end{equation}
where $\mathbb{H}=0$ if the output and the input values are both larger or smaller than the ground-truth value and $\mathbb{H}=1$ if they are not on the same side. More specifically, when there are two equally-good output values, this function prefers the one that is closer to the input and penalizes the other that diverges too much from the input. We find that $\lambda=1.5\sim2.5$ leads to reasonable output sampling maps with sufficient details. 

We also find that an adversarial loss used in previous work \cite{yu2019free, yi2020contextual, bako2019offline} offers only slight improvements on recovering details in the sampling map and is not very helpful for final rendering in most cases, so we avoid using adversarial losses for simplicity.

\Comment{
Inspired by previous image inpainting \cite{yu2019free, yi2020contextual} and path guiding \cite{bako2019offline} work, we also include an adversarial loss using Generative adversarial networks (GANs). Given a large amount of the ground-truth sampling maps, we train a discriminator to classify whether a reconstructed sampling map is real or fake. The adversarial loss for the neural reconstruction network and for the discriminator are:
\begin{equation}
    \mathbb{L}_{\text{adv}} = -\mathbb{E}_{\SMap_{d} \sim P_{g}}[\SMap_{d}]
    \label{eqn:loss_adv}
\end{equation}
\begin{equation}
    \mathbb{L}_{\text{discrim}} = \mathbb{E}_{\SMap_{d} \sim P_{g}}[\SMap_{d}] - \mathbb{E}_{\hat{\SMap_{d}} \sim P_{r}}[\hat{\SMap_{d}}]
    \label{eqn:loss_dis}
\end{equation}

where $P_{g}$ and $P_{r}$ are the distributions of reconstructed and real sampling maps. This part is trained using WGAN \cite{gulrajani2017improved}. The final loss is:
\begin{equation}
    \mathbb{L}_{\text{final}} = \mathbb{L}_{\text{rec}} + \beta \mathbb{L}_{\text{adv}}
    \label{eqn:loss_final}
\end{equation}
where $\beta = 10^{-3}$ is the weight for adversarial loss.
}

\subsection{Discussion}
Our network focuses on reconstructing high-quality sampling functions for local path sampling. 
This is a central sub-problem in many path guiding frameworks.
Note that, this problem of sampling map regression is independent of other sub-modules in path guiding. 
We thus train our network independently without relying on any specific guiding frameworks;
we randomly construct 3D voxels with various sizes in training scenes, and compute sampling maps with both sparse and dense photons to obtain many training pairs (please refer to Sec.~\ref{sec:impl} for details of data generation and training).

Note that, our learning-based sampling map reconstruction module can potentially be applied in many existing path-guiding frameworks (like \cite{jensen1995importance,vorba2014line, muller2017practical, Rath2020}), and improves the traditional sampling distribution reconstruction modules.
In this work, we present a new framework (Sec.~\ref{sec:neuralguiding}) with adaptive spatial partitioning which iteratively builds sampling maps using our neural network in a hierarchical grid for path guiding.

\section{Neural path guiding using a hierarchical grid}
\label{sec:neuralguiding}
In this section, we introduce our novel path guiding framework that leverages our presented deep network to reconstruct high-quality sampling maps in a hierarchical grid.
Our full framework is illustrated in Algorithm~\ref{algo_pseudo}. 
As shown in Algorithm~\ref{algo_pseudo}, we first initialize a grid (Line~1) and then utilize an iterative process (Line~2$\sim$39) to build a hierarchical grid with per-voxel sampling maps for path guiding and rendering.
In each iteration, we trace camera paths (Line~3$\sim$24); these paths can be guided (Line~7$\sim$11) when tracing, and they are used to detect valid voxels (Line~12) and compute the mixture weight of one-sample MIS (Line~17$\sim$20).
We also trace photons (Line~25$\sim$35) per iteration;
in each valid voxel, we accumulate photon energy (Line~29) that is required by our network and also collect other photon statistics for subdividing the hierarchical grid (Line~30$\sim$33).
We then reconstruct the sampling map in each valid voxel using our pre-trained deep neural network at the end of each iteration (Line~36$\sim$38); these sampling maps are used to guide the following path tracing.
After the iterative process, we apply a final path tracing to compute the final output image (Line~40).

Essentially, we iteratively trace camera paths and photons for adaptively partitioning the scene space to a hierarchical grid (see Sec.~\ref{sec:grid} and green blocks in Algorithm~\ref{algo_pseudo}). 
Meanwhile, the photon samples are also used for computing the sampling maps in each voxel (see Sec.~\ref{sec:reconstruction} and blue blocks in Algorithm~\ref{algo_pseudo}) to guide the tracing of the paths in the following iterations; the path samples are also used for rendering and computing the weight $\alpha$ for one-sample MIS (see Sec.~\ref{sec:guiding} and red blocks in Algorithm~\ref{algo_pseudo}).
After a total number of $T$ iterations, we do a final path tracing pass (see Sec.~\ref{sec:finaltracing} and the purple block in Algorithm~\ref{algo_pseudo}) with $\PathF$ spp to render the image. 
The final rendering result is computed from all path samples in the iterations (except for the first iteration that is not guided) and the final pass.
Note that, we double the number of paths and photons after each iteration, so that both the quality of the input raw sampling maps and final rendering can be progressively improved; this leads to $2^t \PathN$ spp paths and $2^t \PhotonN$ photon rays for iteration $t$, where $\PathN$ is the initial spp and $\PhotonN$ is the initial number of photon rays in the first iteration.

\begin{figure}[t]
    \includegraphics[width=0.95\linewidth]{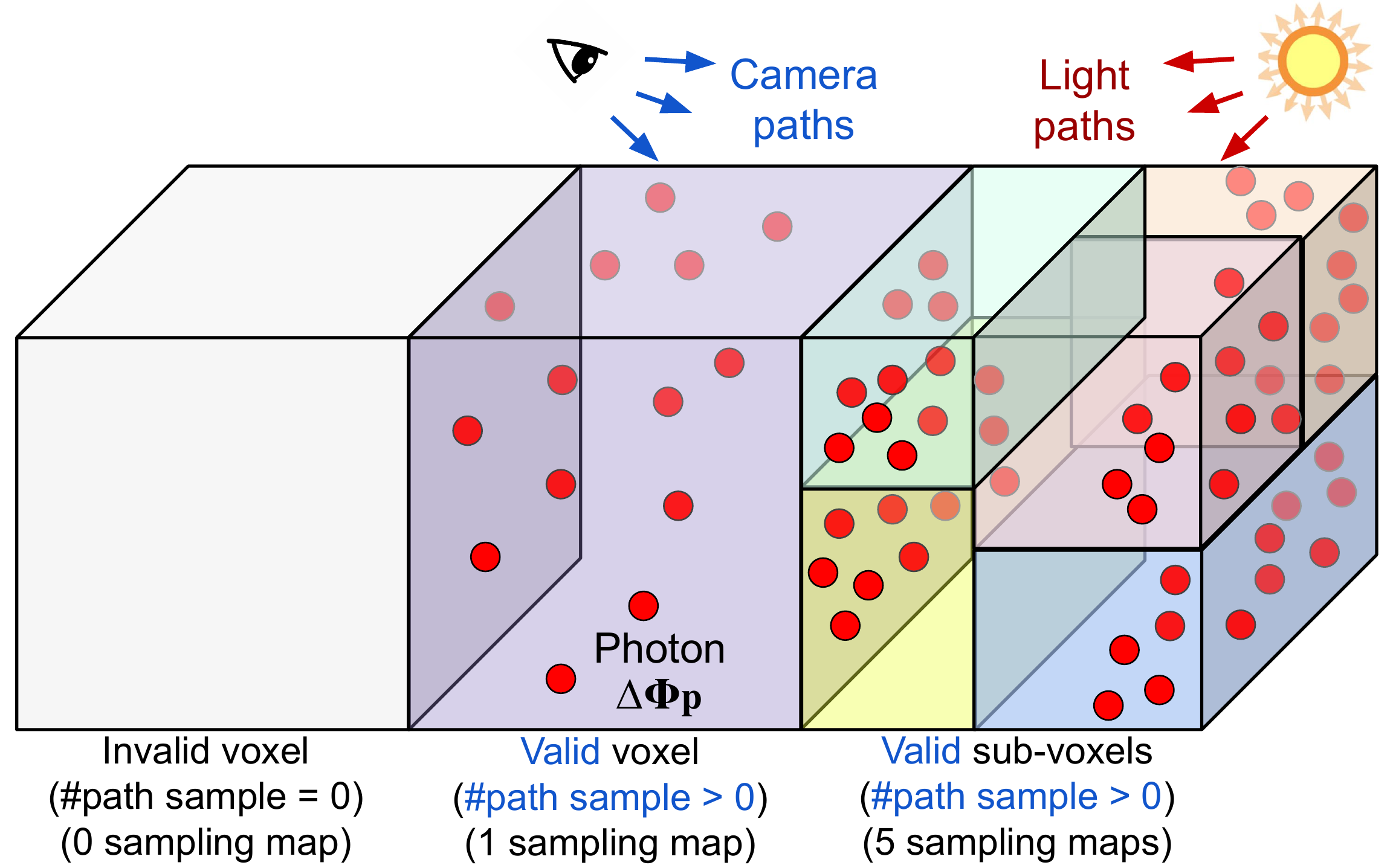}
    \caption{Our proposed hierarchical grid spatial caching structure. The path samples are used to detect valid voxels to store sampling maps. A voxel is subdivided into a binary tree based on the local photon statistics (red points with energy $\Delta\Phi_{p}$). In this example, there are 3 coarse-level voxels in the regular grid (from left to right). The left invalid voxel does not receive any path sample thus no sampling maps or photons are stored. The middle valid voxel stores one sampling map from accumulated photon energies. The right valid voxel gets refined by subdivision and stores 5 sampling maps, one for each sub-voxel.}
    \label{fig:grid_vis}
\end{figure}

\Comment{
\begin{figure}[!ht]
    \includegraphics[width=\linewidth]{image/system.pdf}
    \caption{The proposed path guiding framework. We build a hierarchical grid to store per-voxel sampling maps. In each iteration, we trace camera rays and photons for spatial grid partitioning and sampling map reconstruction. Then, the learned sampling maps are used to importance sampling the remaining path samples. \ZX{I'm not sure if this is useful since we already have the pseduo code} }
    \label{fig:system}
\end{figure}
}

\subsection{An adaptive hierarchical grid for path guiding}
\label{sec:grid}
Since a pure uniform spatial structure (often achieved by spatial cache points in early works \cite{jensen1995importance,vorba2014line}) is very expensive and impractical for large-scale scenes,
recent works often utilize a KD-Tree \cite{muller2017practical} to adaptively partition the space, starting from a single root node that covers the entire scene.
This coarse-to-fine spatial structure is effective, and, in fact, also necessary for many pure online-learning approaches \cite{guo2018primary,Rath2020}, since they need to use a large number of (path) samples that can be only acquired in a large spatial region at an early stage.
In contrast, our deep learning based approach can reconstruct a high-quality sampling map from a sparse set of photons; consequently, starting from a highly coarse spatial partitioning is unnecessary and also even inefficient for our approach.
Therefore, we propose to use a hierarchical grid for spatial partitioning, which combines uniform and adaptive spatial partitioning (Fig.~\ref{fig:grid_vis}).

\paragraph{An initial regular grid.} We start from a regular grid, uniformly dividing the entire scene at a relatively coarse level (see the three coarse voxels in Fig.~\ref{fig:grid_vis}), as the initial spatial structure (Line~1 in Algorithm~\ref{algo_pseudo}); the initial grid is still coarse but relatively much denser than a shallow KD-Tree used in early stages in previous work \cite{muller2017practical}.   
This regular grid enables reconstructing more locally representative sampling maps, leading to good path guiding quality even at the first iteration in our framework, which fully utilizes the benefits of our offline trained deep neural network.
Starting from this regular grid, we iteratively sub-partition each grid voxel into a local KD-tree (Line~30$\sim$33 in Algorithm~\ref{algo_pseudo}), leveraging the statistical information of per-iteration paths and photons; 
this results in a hierarchical grid that adaptively covers the scene and we reconstruct the sampling maps per voxel in each iteration accordingly.

\paragraph{Detecting valid voxels using paths.} While we can compute a sampling map for every voxel in the grid for path guiding, this is usually costly and in fact unnecessary, since many voxels may not be reached by any paths from the viewpoint when rendering a large scene.
Therefore, we leverage the per-iteration camera paths to detect which voxels are necessary for rendering this viewpoint (Line~12 in Algorithm~\ref{algo_pseudo}). 
Specifically, when tracing the $2^t \PathN$ spp path samples in each iteration, we mark a voxel (that hasn't been marked before) as a new valid voxel, if there is at least one bounce point of the paths located in the voxel (see the two valid voxels in Fig.~\ref{fig:grid_vis}).
In other words, we only consider a voxel for sampling map reconstruction and further spatial partitioning when it is known to be necessary (at least likely necessary) in rendering in the following iterations.
This avoids the waste of reconstructing many unnecessary sampling maps and local sub-KD-trees. 
Once a voxel is marked as valid, we start accumulating photons in the voxel for sampling map reconstruction and further subdivision of the voxel.

\paragraph{Voxel subdivision.}
It is not efficient to use a regular grid for spatial partitioning, since various local spatial regions may involve highly diverse geometry, appearance, and lighting distributions.
Therefore, we iteratively subdivide the initial regular grid into a hierarchical grid, where a voxel is divided into a binary tree similar to a local KD-tree if necessary (Line~30$\sim$33 in Algorithm~\ref{algo_pseudo}). 
Our hierarchical grid is built to adapt to the complexity of local geometry and incident light fields. In the very beginning iterations, we trace small numbers of light paths and photon data is temporarily stored in each voxel. 
We leverage the statistics of accumulated photons in the current iteration in each valid voxel for the voxel's possible subdivision. 
In particular, for each valid voxel $j$, we consider $\PhotonVoxelM_j$ -- the total number of photons hitting the voxel through iterations -- and $\PhotonNormalVar_j$ -- the variance of the surface normals at the photon hitpoints.
A voxel is split into two sub-voxels by the middle of photon positions along an axis (just like KD-tree construction), if $\PhotonVoxelM_j > \PhotonVoxelM_{\text{thr}}$ or $\PhotonNormalVar_j > \PhotonNormalVar_{\text{thr}}$, where $\PhotonVoxelM_{\text{thr}}$ and $\PhotonNormalVar_{\text{thr}}$ are two predefined thresholds and we recursively apply our
subdivision criterion to sub-voxels (see the right voxel in Fig.~\ref{fig:grid_vis}). 
Once a voxel is subdivided, its two sub-voxels are kept as valid, accumulating photons from the current iteration and waiting for photons in following iterations to reconstruct sampling maps. 
These simple photon statistics are easy to compute, enabling efficient subdivision. This photon-based subdivision process subdivides voxels that either have complex light fields (dense photons) or complex geometry (large normal variation). 
Our method allows these complex voxels to utilize more local and accurate sampling maps in the following iterations, thus leading to more accurate renderings.


\subsection{Sampling map reconstruction}
\label{sec:reconstruction}
Apart from determining the subdivision in the hierarchical grid, the main goal of tracing the per-iteration photons is to reconstruct the per voxel sampling maps for path guiding.
For any valid voxel (marked by camera paths), we accumulate photon energies to compute the energy map of the voxel (Line~29 in Algorithm~\ref{algo_pseudo}), as is expressed by Eqn.~\ref{eqn:photon}. 
The energy map records the sum of the energies of all hitting photons $\Delta\Phi_{p}$ in the voxel through the current and all previous iterations. 
The per-pixel accumulated energy $\Phi_{j,k}$ in an energy map will be normalized, which leads to a raw sampling map $\SMap_{e,t}$ that is sent as input to the network to reconstruct the sampling map in iteration $t$.
As discussed in Sec.~\ref{sec:learning}, we also provide additional input buffers (photon count, previous raw sampling map, and binary mask) for the network. 
Specifically, we record the number of accumulated photons and also keep the raw sampling map and number of photons in the previous iteration to construct the network input.
After tracing all photons in an iteration, we collect all valid voxels that have new photons arrived and reconstruct the sampling maps $\SMap_d$ using our deep neural network for path guiding (Line~37 in Algorithm~\ref{algo_pseudo}). 
As mentioned, we exponentially increase the photon count per iteration with a base of 2, similar to the growth of path samples by \citet{muller2017practical}, so that the number of photons consumed by the input sampling map is roughly doubled after each iteration. 
Once a sampling map is reconstructed at a voxel in one iteration, the map is used in the following iterations and the final path tracing pass to guide the path sampling in the voxel.


\subsection{Path guiding and one-sample MIS}
\label{sec:guiding}
In any iteration, if a path hits a voxel that doesn't have a sampling map, we just use standard BSDF sampling at the bounce point (Line~8 in Algorithm~\ref{algo_pseudo}); such a voxel is usually still an invalid voxel, which will be marked as valid and start accumulating photons immediately in the same iteration, allowing for path guiding in the following iterations.
On the other hand, once a path ray hits a valid voxel that has a reconstructed sampling map, path guiding can be achieved by doing importance sampling on the sampling map (where a CDF is built via a fast cumulative sum over pixels on GPUs, just like sampling an environment map).
Since our sampling map only considers the incident radiance (and a cosine term), we apply a one-sample MIS similar to previous works to combine guided sampling and BSDF sampling (Line~10 in Algorithm~\ref{algo_pseudo}), as discussed in Eqn.~\ref{eqn:target_mis}.
\Comment{
Once the hierarchical grid with per-voxel sampling maps is iteratively constructed,
we proceed a final path tracing to render the final image, guided by the reconstructed sampling maps. Each sampling map is accessed efficiently in $O(1)$ time using the hashing index in the grid. If a sub-voxel is invalid or if the photons are too sparse, the reconstructed sampling map in its parent voxel is used for guiding.
Since we only consider the incident radiance (integrated as energy) for our sampling maps, we apply a one-sample MIS similar to previous works to combine guided sampling and BSDF sampling, as discussed in Eqn.~\ref{eqn:target_mis}.
}
The combined sampling strategy however requires a parameter $\alpha$ that determines how often either sample strategy is selected.
Usually, $\alpha=0.5$ is a simple choice and performs reasonably well in previous work \cite{muller2017practical}. 
An $\alpha$ that is learned via online optimization (\cite{mueller19guiding}) is also presented for better performance but requires expensive online training.

We present a heuristic $\alpha$ computation technique, based on path statistics (Line~17$\sim$20 in Algorithm~\ref{algo_pseudo}); though simple, it results in effective per-voxel $\alpha_j$ in practice for high-quality path guiding.
In particular, we initially use $\alpha_j=0.5$ in each voxel. 
While tracing paths in each iteration, we first construct all paths using one-simple MIS according to the current per-voxel mixture weights $\alpha_j$. 
And once a full path is constructed, we compute the actual sub-path contribution (often known as throughput, Line~17 in Algorithm~\ref{algo_pseudo}) for every bounce point $b$ on the path as $\nu_b = L(\Px_b, \omega_b) \cos \theta_b f_{r}(\Px, \omega_b, \DirO)$, where $\Px_b$ is the position of the bounce point, $\omega_b$ is the sampled direction (that can come from either BSDF or guided sampling), $L(\Px_b, \omega_b)$ is computed by consecutively multiplying the light radiance, BSDFs, and inversed sampling PDFs through all following bounce points as in a standard Monte Carlo path sample.
Meanwhile, for each voxel $j$, we accumulate all bounce contributions $\nu_b$ (of the bounces that are in the voxel, i.e., $\Px_b \in \DeltaA_j$)
in $\nu_{j,\text{B}}$ and $\nu_{j,\text{G}}$, according to from which distribution $\omega_b$ is sampled (Line~18 in Algorithm~\ref{algo_pseudo}).
Specifically, $\nu_{j,\text{B}}$ records the sum of all path contributions $\nu_b$ if its direction $\omega_b$ is sampled by BSDF sampling, and $\nu_{j,\text{G}}$ records the sum of $\nu_b$ if $\omega_b$ is sampled by guided sampling. 
We also record the numbers of bounces sampled by the two sampling strategies as $Q_{j, \text{B}}$ and $Q_{j, \text{G}}$ (Line~19 in Algorithm~\ref{algo_pseudo}) in each valid voxel. 
Once $\PathVoxelM_{j,\text{B}} \geq 50$ and $\PathVoxelM_{j,\text{G}} \geq 50$ sub-paths are sampled in a valid voxel $j$ (Line~20 in Algorithm~\ref{algo_pseudo}), we use the ratio of the averaged $\nu_{j,\text{B}}$ and $\nu_{j,\text{G}}$ to determine the mixing weight $\alpha_j$ for following path guiding:

\begin{equation}
\alpha_j = \frac{\overline{\nu}_{j,\text{B}}}{\overline{\nu}_{j,\text{B}} + \overline{\nu}_{j,\text{G}}},
\label{eqn:mis_alpha}
\end{equation}
where $\overline{\nu}_{j,\text{B}} = \nu_{j,\text{B}}/Q_{j,\text{B}}$ and $\overline{\nu}_{j,\text{G}} = \nu_{j,\text{G}}/Q_{j,\text{G}}$.
Correspondingly, our one-sample MIS is expressed by:
\begin{equation}
    \small
    \begin{split}
    p_\text{MIS}(\DirI) & = \frac{\overline{\nu}_{j,\text{B}}}{\overline{\nu}_{j,\text{B}} + \overline{\nu}_{j,\text{G}}} p_{\text{BSDF}}(\DirI) + \frac{\overline{\nu}_{j,\text{G}}}{\overline{\nu}_{j,\text{B}} + \overline{\nu}_{j,\text{G}}}p_{\text{guide}}(\DirI).
    \end{split}
    \label{eqn:mis_coef}
\end{equation}
\Comment{
\begin{equation}
    \small
    \begin{split}
    p_\text{MIS}(\DirI) & = \frac{\nu_{j,\text{B},\text{avg}}}{\nu_{j,\text{B},\text{avg}} + \nu_{j,\text{G},\text{avg}}} p_{\text{BSDF}}(\DirI) + \frac{\nu_{j,\text{G},\text{avg}}}{\nu_{j,\text{B},\text{avg}} + \nu_{j,\text{G},\text{avg}}}p_{\text{guide}}(\DirI).
    \end{split}
    \label{eqn:mis_coef}
\end{equation}
}
We set $\alpha_j=1.0$ if BSDF is a delta function and clamp $\alpha_j$ between 0.2 and 0.8 otherwise. This heuristic mixing weight considers the data that reflects the actual performance of BSDF sampling and guiding sampling, leading to effective one-sample MIS sampling in our path guiding.

\begin{figure}[h]
    \includegraphics[width=\linewidth]{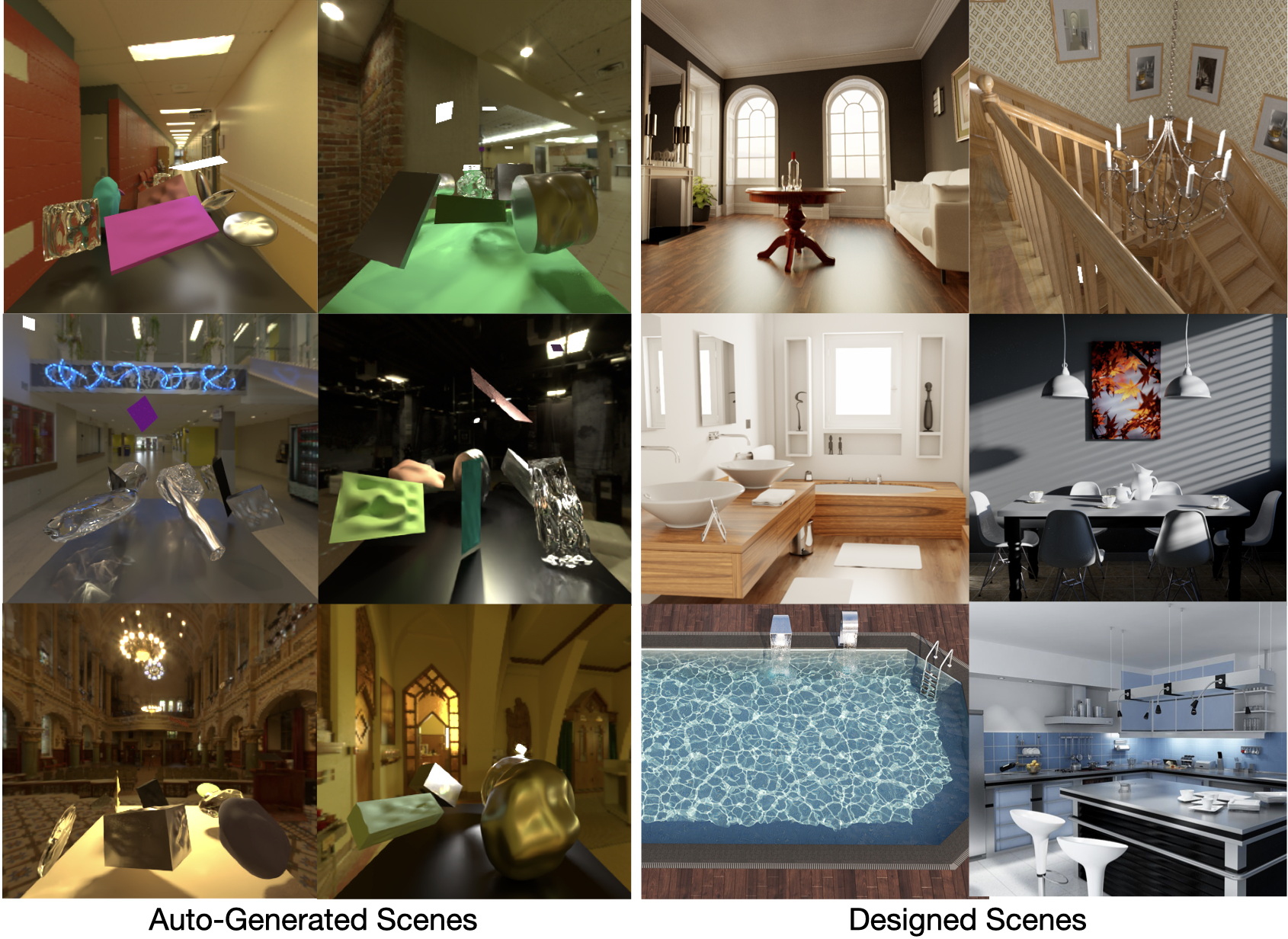}
    \caption{Example scenes used for training our proposed neural network.}
    \label{fig:scenes}
\end{figure}

\subsection{Rendering and final path tracing}
\label{sec:finaltracing}
Our learning based approach is able to reconstruct high-quality sampling maps from very sparse photons, leading to efficient guided path tracing even in early iterations.
The first iteration paths are not guided at all since there are no sampling maps reconstructed yet.
However, thanks to our deep neural network, our path guiding is often of very good quality starting from the second iteration. 
We therefore leverage all path sampling starting from the second iteration for rendering the final image.

While we can keep iteratively tracing more photons and refining our sampling maps, we find that our reconstructed sampling maps are often of very high quality after $T=$ 4$\sim$9 iterations. As a result, continuing tracing more photons afterwards merely leads to marginal sampling improvement.
Therefore, we choose to stop tracing photons after $T=$ 4$\sim$9 iterations, fix the per-voxel sampling maps, and switch to do pure path tracing guided by the fixed sampling maps using a number of $\PathF$ spp as needed.
This is called the final path tracing pass in our framework (Line~40 in Algorithm~\ref{algo_pseudo}).
Our final rendered image is computed from all path samples traced in all $T$ iterations and the final path tracing pass.

\section{Implementation}
\label{sec:impl}

\paragraph{Dataset generation and neural network training.}
We create a large scale dataset to train our sampling map reconstruction network. Our dataset consists of both designed scenes and auto-generated scenes as shown in Fig.~\ref{fig:scenes}.
We first collect available online scenes designed by researchers or artists, by collecting several released scenes from previous work and purchasing scenes from several websites \cite{resources16, Mitsuba, evermotion2012evermotion, trader4cg, squid3d, blend2016blend}.
This leads to 32 designed scenes in total, including multiple realistic indoor and outdoor scenes; we use 20 from them in our training set and the rest for testing our algorithm.
To enhance the generalizability of our network, we further enlarge our training set by auto-generating many more scenes.
In particular, we procedurally generate 500 scenes using randomized shape primitives, materials, and area lights, similar to \cite{zhu2020deep,xu2018deep}. 
We also leverage a complex lighting dataset \cite{gardner2017learning} and randomly select an environment map for each generated scene as its additional illumination.
This auto-generation process largely increases the diversity and complexity of our training scenes, leading to better generalization on novel testing scenes. 

\newcolumntype{P}[1]{>{\centering\arraybackslash}p{#1}}
\newcolumntype{M}[1]{>{\centering\arraybackslash}m{#1}}

\newcommand{\ScnCE}{\textsc{Caustics Egg}}
\newcommand{\ScnVA}{\textsc{Veach Ajar}}
\newcommand{\ScnBR}{\textsc{Bathroom}}
\newcommand{\ScnH}{\textsc{Hotel}}
\newcommand{\ScnSC}{\textsc{Staircase}}
\newcommand{\ScnLR}{\textsc{Living Room}}
\newcommand{\ScnSS}{\textsc{Spaceship}}
\newcommand{\ScnCR}{\textsc{Classroom}}
\newcommand{\ScnWC}{\textsc{Wild Creek}}
\newcommand{\ScnT}{\textsc{Torus}}
\newcommand{\ScnK}{\textsc{Kitchen}}
\newcommand{\ScnP}{\textsc{Pool}}

\begin{table}[h]
	\begin{center}
		
		\begin{tabular}{p{0.7in}|p{0.37in}|P{0.42in}|P{0.42in}|P{0.37in}|P{0.25in}}  
			\hline  
			Component & \centering Path ($\%$) & Photon ($\%$) & Neural rec ($\%$) & Path ($\%$) & Time (min) \\
			\hline
			Algorithm~\ref{algo_pseudo} & \centering LN~3$\sim$24 & LN~25$\sim$35 & LN~36$\sim$38 & LN~40 & / \\
			\hline
			Device & \centering CPU & CPU & GPU & CPU & / \\
			\hline
			Phase & \multicolumn{3}{c|}{iterative process (when $t<T$)} & final & / \\
			\hline
			\textsc{Caustics Egg} &\centering 13.91 & 21.83 & 8.36  & 55.88 & 4.0 \\
			\hline
			\textsc{Veach Ajar} &\centering 14.58 & 21.08 & 5.76  & 58.56 & 18.0 \\
			\hline
			\textsc{Bathroom} &\centering 15.42 & 11.86 & 9.77  & 62.93 & 5.0 \\
			\hline
			\textsc{Hotel} &\centering 15.05 & 18.58 & 5.89  & 60.46 & 20.0 \\
			\hline
			\textsc{Staircase} &\centering 15.79 & 15.77 & 5.01  & 63.41 & 11.0 \\
			\hline
			\textsc{Living Room} &\centering 16.42 & 11.73 & 5.89  & 65.94 & 11.0 \\
			\hline
			\textsc{Spaceship} &\centering 16.49 & 9.20 & 8.06  & 66.23 & 3.0 \\
			\hline
			\textsc{Classroom} &\centering 15.33 & 16.68 & 6.39  & 61.57 & 13.0 \\
			\hline
			\textsc{Wild Creek} &\centering 17.05 & 8.41 & 6.02  & 68.50 & 10.0 \\
			\hline
			\textsc{Torus} &\centering 15.72 & 12.78 & 8.32  & 63.16 & 4.0 \\
			\hline
			\textsc{Kitchen} &\centering 14.35 & 19.06 & 8.94  & 57.63 & 4.0 \\
			\hline
			\textsc{Pool} &\centering 16.78 & 8.22 & 7.57  & 67.40 & 4.0 \\
			\hline
		\end{tabular}
		\caption{Running time. Percentages of running time of different components in the proposed system are shown in the table for different testing scenes. The total rendering time for each scene is also shown in the rightmost column. The time distribution varies depending on the scene complexity and light setup.}
		\label{tab:runtime}
	\end{center}
\end{table}

\begin{table*}[t]
	\begin{center}
		
		\begin{tabular}{p{0.7in}|P{0.3in}P{0.4in}P{0.4in}P{0.4in}P{0.4in}P{0.4in}|P{0.3in}P{0.4in}P{0.4in}P{0.4in}P{0.4in}P{0.4in} }  
			\hline  
			Scene/Method & PT & \cite{bako2019offline}  & \cite{vorba2014line}  &  \cite{muller2017practical} 
			 &  \cite{Rath2020} & \textbf{Ours} &  PT & \cite{bako2019offline}  & \cite{vorba2014line} & \cite{muller2017practical} & \cite{Rath2020} & \textbf{Ours} \\
			 \hline
			 Metric & \multicolumn{6}{c|}{rMSE $\downarrow$} & \multicolumn{6}{c}{SSIM $\uparrow$}\\
			\hline
			\multirow{1}{*}{\textsc{Caustics Egg}}& 0.3187 & 0.1353  & 0.0462 & \cellcolor{yellow!50} 0.0311   & \cellcolor{orange!30} 0.0121 & \cellcolor{pink} 0.0052 & 0.1017  & 0.1824  & 0.3472  & \cellcolor{yellow!50} 0.4581 & \cellcolor{orange!30} 0.7006 & \cellcolor{pink} 0.8242 \\
			\hline
			\multirow{1}{*}{\textsc{Veach Ajar}}& 0.3684 & 0.2585  & 0.0154 & \cellcolor{yellow!50} 0.0073   & \cellcolor{orange!30} 0.0047 & \cellcolor{pink} 0.0011 & 0.0474  & 0.0898   & 0.4579  & \cellcolor{yellow!50} 0.5455 & \cellcolor{orange!30} 0.6325 & \cellcolor{pink} 0.8572 \\
			\hline
			\multirow{1}{*}{\textsc{Bathroom}}& 0.0610 & 0.0403  & \cellcolor{yellow!50} 0.0204  & 0.0249   & \cellcolor{orange!30} 0.0142 & \cellcolor{pink} 0.0050 & 0.4481  & 0.4725   & \cellcolor{yellow!50} 0.5472  & 0.5260 & \cellcolor{orange!30} 0.5924 & \cellcolor{pink} 0.7427 \\
			\hline
			\multirow{1}{*}{\textsc{Hotel}}& 0.4176 & 0.2607 & 0.2838  & \cellcolor{yellow!50} 0.0812   & \cellcolor{orange!30} 0.0792 & \cellcolor{pink} 0.0276 & 0.0695 & 0.1155  & 0.0914  & \cellcolor{yellow!50} 0.2665 & \cellcolor{orange!30} 0.2801 &  \cellcolor{pink} 0.4378 \\
			\hline
			\multirow{1}{*}{\textsc{Staircase}}& 0.0176 & 0.0183  & 0.0110 &\cellcolor{yellow!50} 0.0045   & \cellcolor{orange!30} 0.0038 &  \cellcolor{pink} 0.0013 & 0.4810  & 0.4957   & 0.6513  & \cellcolor{yellow!50} 0.7337 & \cellcolor{orange!30} 0.8626 & \cellcolor{pink} 0.8951 \\
			\hline
			\multirow{1}{*}{\textsc{Living Room}}& 0.1928 & 0.1553  &  \cellcolor{orange!30} 0.0235 & 0.0468   & \cellcolor{yellow!50} 0.0416 & \cellcolor{pink} 0.0060 & 0.1360  & 0.1719 & \cellcolor{orange!30} 0.4734  & 0.2960 & \cellcolor{yellow!50} 0.3327 & \cellcolor{pink} 0.6576 \\
			\hline
			\multirow{1}{*}{\textsc{Spaceship}}& 0.2212 & 0.0914  & \cellcolor{orange!30} 0.0198  & 0.0716   & \cellcolor{yellow!50} 0.0389 & \cellcolor{pink} 0.0137  & 0.5610  & 0.7476 & \cellcolor{orange!30} 0.8611   & 0.7452 & \cellcolor{yellow!50} 0.8124 & \cellcolor{pink} 0.8793 \\
			\hline
			\multirow{1}{*}{\textsc{Classroom}}& 0.0733 & 0.0514  & 0.0124 & \cellcolor{yellow!50} 0.0085   & \cellcolor{orange!30} 0.0038 & \cellcolor{pink} 0.0021   & 0.2789  & 0.3037 & 0.5756   & \cellcolor{yellow!50} 0.6352 & \cellcolor{orange!30} 0.7681 & \cellcolor{pink} 0.8234 \\
			\hline
			\multirow{1}{*}{\textsc{Wild Creek}}& 0.1425 & 0.1100  & \cellcolor{yellow!50} 0.0560 & 0.0618   & \cellcolor{orange!30} 0.0549 & \cellcolor{pink}  0.0382  & 0.3023  & 0.3734 & \cellcolor{yellow!50} 0.4890  & 0.4852 & \cellcolor{orange!30} 0.5386 & \cellcolor{pink} 0.6222 \\
			\hline
			\multirow{1}{*}{\textsc{Torus}}& 0.0511 & 0.0425 & \cellcolor{yellow!50} 0.0150 & \cellcolor{orange!30} 0.0015   & \cellcolor{orange!30} 0.0015 & \cellcolor{pink} 0.0005  & 0.2610  & 0.6660 & 0.7864  & \cellcolor{yellow!50} 0.9150 & \cellcolor{orange!30} 0.9300 & \cellcolor{pink} 0.9529 \\
			\hline
			\multirow{1}{*}{\textsc{Kitchen}}& 0.0644 & 0.0578  & 0.0249 & \cellcolor{yellow!50} 0.0063   & \cellcolor{orange!30} 0.0035 &  \cellcolor{pink} 0.0030  & 0.3898  & 0.4173  & 0.4655  & \cellcolor{yellow!50} 0.6753 & \cellcolor{orange!30} 0.7873 & \cellcolor{pink} 0.8168 \\
			\hline
			\multirow{1}{*}{\textsc{Pool}}& 0.1175 & 0.0528  & 0.0026  & \cellcolor{yellow!50} 0.0025   & \cellcolor{orange!30} 0.0016 &  \cellcolor{pink} 0.0011  & 0.2264  & 0.4595 & 0.8551  &  \cellcolor{yellow!50} 0.8598 & \cellcolor{orange!30} 0.9364 & \cellcolor{pink} 0.9510 \\
			\hline
		\end{tabular}
		
		\caption{Quantitative comparison. We compare our results and the results of \cite{bako2019offline,vorba2014line,muller2017practical,Rath2020} with equal rendering time. We show the corresponding rMSEs and SSIMs of the rendered full images of the 12 testing scenes. Red, orange, and yellow denote the best, the second-best, and the third-best method in terms of rMSE (lower is better) and SSIM (higher is better). Our method achieves the best results on all testing scenes. The total rendering time for each scene is presented in Tab.~\ref{tab:runtime}.}
		\vspace{-5mm}
		\label{tab:quant_rmse}
	\end{center}
\end{table*}

We reconstruct sampling maps with the same resolution of $128\times64$.
As expected, if memory allows, a higher resolution of sampling maps often leads to better rendering quality. 
While our rendering quality degrades with a lower resolution, we find that, even using $64\times32$ sampling maps, our method can still outperform previous state-of-the-art methods (see Fig.~\ref{fig:ablation}).
Our network aims to reconstruct a sampling map of a local 3D voxel. 
We partition the space of each training scene uniformly using a regular grid with a random resolution ranging from $50^3$ to $200^3$. 
This makes our network generalize to various voxel sizes, naturally enabling high-quality sampling map reconstruction for any voxel at any depth in a hierarchical grid.
To further augment the data, we also randomly rotate the world coordinate frame when partitioning.
We trace photons in each scene and compute sampling maps based on Eqn.~\ref{eqn:photon} using both sparse and dense photons, which constructs the input and ground-truth training pairs. 
The total number of training pairs in our dataset is about 10.5 million.
To make the network generalize well on different iterations in our path guiding,
for each training scene, we randomly select an iteration number $t$ from 1 to 12, and compute the corresponding input sampling map using the photons generated by the $2^t \PhotonN$ light paths. On the other hand, the ground-truth sampling map of each voxel is computed by accumulating photons generated through 20 iterations for each scene. 

During rendering, the number of photons in different voxels can be highly different (from several to several thousand), leading to highly diverse input distributions;
we therefore train multiple networks as a mixture of experts\cite{jacobs1991adaptive}, and make each network focus on a certain range of input photon numbers in a voxel. 
Specifically, we train five networks separately and the corresponding ranges of photon numbers are $[0,100)$, $[100,500)$, $[500,1000)$, $[1000,5000)$, $[5000,\infty)$.
This enables better reconstruction quality compared to using a single network for all cases.
And since our networks are very compact (several MBs), using five different networks does not lead to any memory issues.
We implement our networks using PyTorch.
During training, we use mini-batches with a size of 50 and train each network using ADAM \cite{kingma2014adam} with a learning rate of $1.0 \times 10^{-4}$.
Our network generally converges to a very good optimum after 500K epochs, taking about a week using 8 Nvidia RTX 2080Ti GPUs.

\begin{figure*}[t]
    \includegraphics[width=\linewidth]{image/transparent.pdf}
    \caption{Qualitative and quantitative comparison with equal rendering time. These scenes contain many transparent surfaces and involve complex specular-diffuse interactions; Photon-based methods have a natural advantage over path samples in this case. 
    We show zoomed-in crops with rMSEs in the figure and compare with the results of \cite{vorba2014line}, \cite{muller2017practical}, \cite{bako2019offline} and \cite{Rath2020}. Corresponding equal rendering time for each scene is also listed. Our method achieves the best visual quality and the lowest rMSEs in these challenging cases.}
    \label{fig:transparent}
\end{figure*}

\begin{figure*}[t]
    \includegraphics[width=\linewidth]{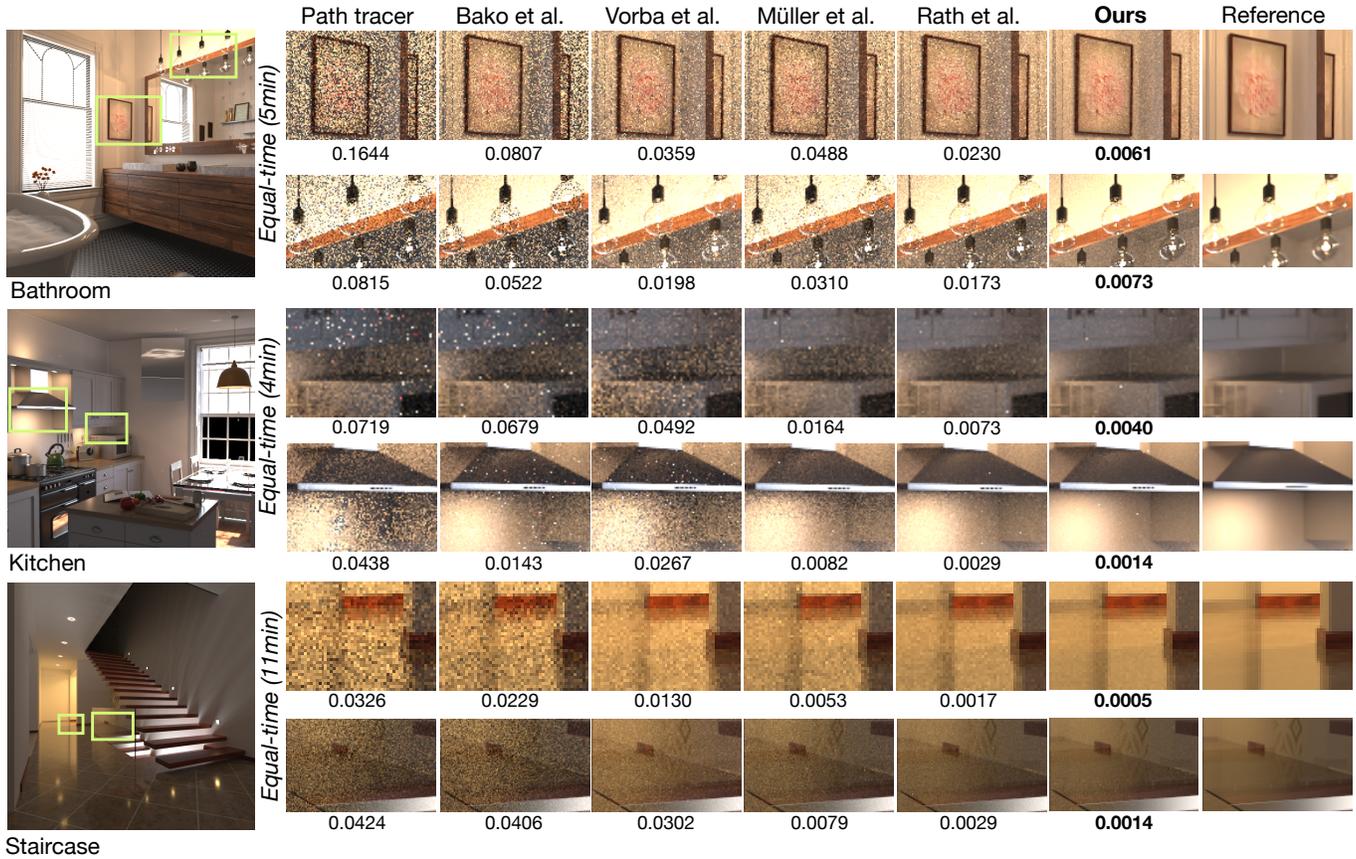}
    \caption{Qualitative and quantitative comparison with equal rendering time. These scenes contain complex indoor lighting, lit by $5\sim10$ area light sources with different shapes. 
    Our deep learning based approach enables accurate sampling map reconstruction for the complex direct and indirect lighting, leading to efficient rendering.
    We show zoomed-in crops with rMSEs in the figure. Corresponding equal rendering time for each scene is also listed. Our method achieves the best visual quality and the lowest rMSEs in these challenging cases.
    }
    \label{fig:complex_light}
\end{figure*}

\begin{figure*}[t]
    \includegraphics[width=\linewidth]{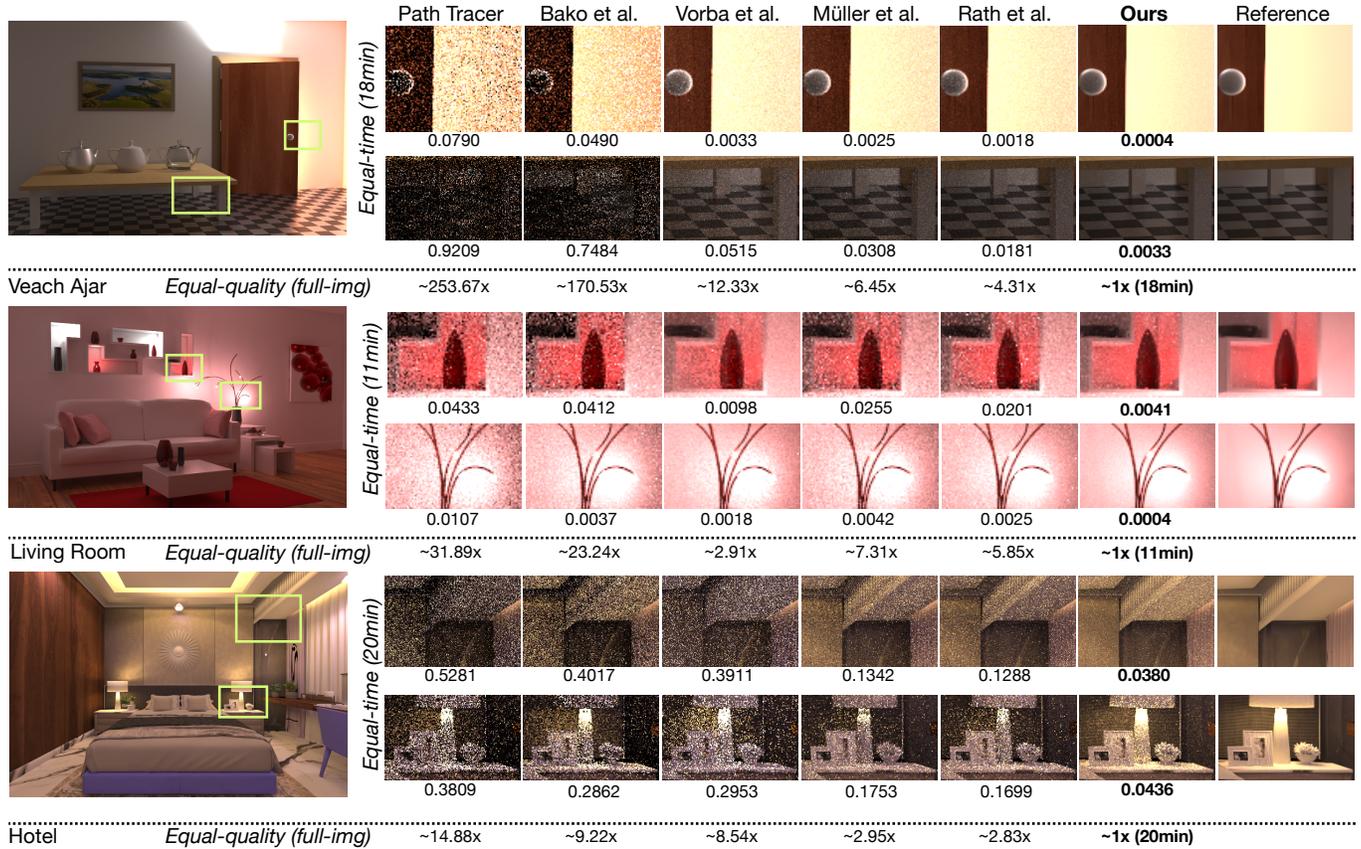}
    \caption{Equal-time and equal-quality comparison.
    Similar to Fig.~\ref{fig:transparent} and \ref{fig:complex_light}, we do qualitative and quantitative equal-time comparisons on crops of the final renderings on these challenging scenes. 
    Our method achieves better qualitative and quantitative results given equal rendering time.
    In addition, we also show equal-quality rendering time comparison. In particular, we list the corresponding rendering time (expressed by the scale to our time) of each method for achieving the same rMSE (that our method achieves in the equal-time comparison, shown in Tab.~\ref{tab:quant_rmse}) of the full image.
    Note that, our method takes significantly less time; the fastest comparison method still requires more than two times the rendering time as our method for each scene.
    }
    \label{fig:budget}
\end{figure*}

\begin{figure}[t]
    \includegraphics[width=\linewidth]{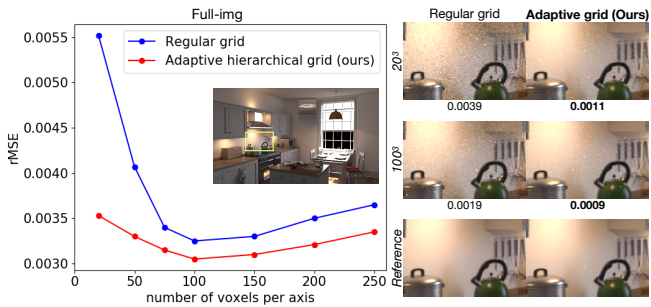}
    \caption{The effect of initial grid resolution and our hierarchical spatial partitioning framework. Ideally, the voxel size should be small enough to reflect the locality of the incident radiance, and large enough to contain enough photons for sampling map reconstruction. In the extreme case, under-partitioning and over-partitioning will both hurt the performance.}
    \label{fig:grid}
\end{figure}

\paragraph{Path guiding details.}
We use $\PathN = 1$ (spp) for all our experiments, leading to $2^t\PathN=2^t$ spp paths for iteration $t$.
We also correspondingly trace the same number of light paths ($\PhotonN$ thus equals to the number of pixels) per iteration for distributing photons.
The initial regular grid is implemented as a hash grid that can be accessed in $O(1)$ time. Each sub binary tree is like a local KD-tree that can be accessed in $O(\log(n))$ time. Our final hierarchical grid is a hybrid spatial structure and can thus be quickly accessed at rendering time, enabling highly efficient path guiding.
Since our spatial structure is adaptively constructed, our method is not very sensitive to the resolution of the initial grid, and we use a resolution of 
$100^3$ for all the testing scenes. 
For voxel subdivision, we use an iteration-dependent threshold for the photon count, given by $\PhotonVoxelM_{\text{thr}} = c \cdot \sqrt{2^{t}}$ similar to \cite{muller2017practical}, where $t$ is the iteration number and $c$ is a scalar parameter. We find $c = 400 \sim 800$ performs similarly in practice, and we use $c=500$ for all our testing experiments. The normal variance threshold is set to $\PhotonNormalVar_{\text{thr}} = 0.5$.
We also set the maximum depth of a local KD-tree to 8, which already corresponds to a very fine grid and avoids unnecessarily detailed subdivision. In practice, we also only proceed to voxel subdivision in the first two iterations, which results in a reasonable hierarchical grid. 

We use a high-end machine with Intel Core i9-7960X CPU and \Comment{four} Nvidia Titan RTX GPUs for rendering our testing scenes. 
Our framework is implemented in the standard rendering engine Mitsuba \cite{Mitsuba}, and we use the PyTorch C++ API \cite{NEURIPS2019_9015} at rendering time for sampling map reconstruction on GPUs. 
This Mitsuba based implementation ensures a fair comparison with previous methods, most of which are also implemented with Mitsuba. 
In particular, we only use GPUs to do network inference for sampling map reconstruction, while all other parts of the algorithm (including path tracing, photon tracing, ray sampling, radiance computation, spatial grid construction, etc.) are done on the CPU as in the standard Mitsuba renderer.
The CPU and the GPU parts are run in a sequence in our implementation.
We believe this is a fair enough setting when comparing with traditional pure CPU-implemented path guiding methods that do not use neural networks.
In fact, our GPU computation time is only about $10\%$ of the total running time; please refer to Tab.~\ref{tab:runtime} for detailed running times for each of our testing scenes.
In the future, a more efficient implementation in practice can be done by making the GPU part run in parallel with the CPU part or even implementing a pure GPU-based framework leveraging hardware ray tracing in modern GPUs \cite{parker2010optix}.

\section{Evaluation}
\label{sec:eval}

We now present extensive experiments to evaluate our path guiding approach. We first evaluate the rendering quality of our method by comparing against various state-of-the-art path guiding methods quantitatively and qualitatively. 
We then investigate sub-components in our system to justify their effectiveness. Some additional evaluation results can be found in the supplementary material.

\begin{figure*}[t]
    \includegraphics[width=\linewidth]{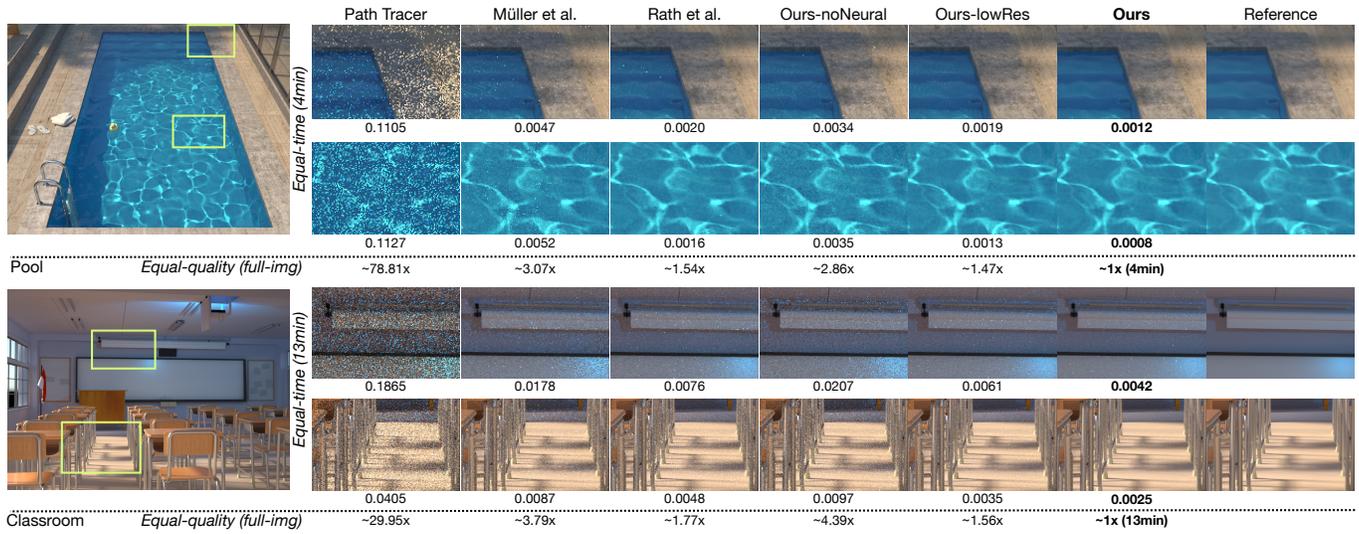}
    \caption{We study the effectiveness of the neural reconstruction module. We compare our full model with a version without neural sampling map reconstruction and a version that uses a lower resolution ($64\times 32$) of sampling maps. We also compare with \cite{muller2017practical} and \cite{Rath2020} on these results. We show crops with rMSEs of the rendered images for each method given equal rendering time. The corresponding equal-quality rendering time to achieve our full-image rMSE is also listed. Our final model achieves the best results among all comparison methods.}
    \label{fig:ablation}
\end{figure*}

\begin{figure}[t]
    \includegraphics[width=0.95\linewidth]{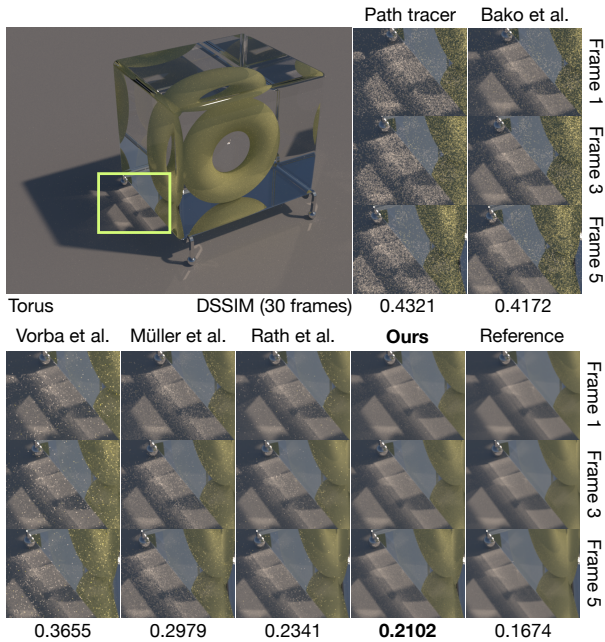}
    \caption{Average DSSIM (lower is better) computed from 30 consecutive frames when the camera is moving fast along a direction. The dissimilarity is affected by both the content change as well as the noise level.}
    \label{fig:temporal}
\end{figure}

\paragraph{Configuration.}
We evaluate our method comprehensively on 12 realistic testing scenes;
the corresponding images of these scenes can be found in Fig.~\ref{fig:transparent}, \ref{fig:complex_light}, \ref{fig:budget}, \ref{fig:ablation} and \ref{fig:temporal}.
These testing scenes include challenging indoor and outdoor cases with complex global illumination, covering a wide range of scene complexity and diversity. Each scene contains both direct and indirect illumination. 
For indoor scenes with outside environment map illumination, we provide the window geometry for sampling light paths from the environment map, facilitating the photon tracing process in these scenes.
For our method, the required time to achieve good rendering quality ranges from 3 to 20 minutes (depending on scene complexity) on these testing scenes.
We demonstrate equal-time comparisons by comparing with four state-of-the-art path guiding methods \cite{bako2019offline,muller2017practical,vorba2014line,Rath2020} on all testing scenes; we also show the corresponding equal-quality rendering time on a few scenes.
In the comparisons, we directly use the released source code of \cite{muller2017practical}, \cite{Rath2020}, and \cite{vorba2014line}, which are all implemented with Mitsuba \cite{Mitsuba} that runs on CPU.
Since there's no publicly available source code of \cite{bako2019offline}, we use our own implementation of it with Mitsuba for all experiments. 
As discussed in Sec.~\ref{sec:impl}, we implement our method also in Mitsuba, mostly running on CPU for fair comparisons, while only the network inference for sampling map reconstruction runs on GPU, which only takes about $10\%$ of the total running time (see Tab.~\ref{tab:runtime} for detailed timing).
Our implementation of \cite{bako2019offline} follows similar CPU and GPU separation, where we run their sampling map reconstruction network on GPUs and run other parts of the algorithm on CPU. 
All comparisons are run on the same machine with the same CPU and GPUs (if needed).
\Comment{\MM{For the equal time comparisons, are all techniques making use of the 4 GPUs?  That wasn't clear in the text and if the neural network method is, and the others (standard path tracing, etc.) aren't then that isn't a completely fair comparison.  Standard path tracing could fire a lot of rays with 4 GPUs - that's a lot of computational horsepower.  Either way, it should be clarified in the text.}}
To better illustrate the effectiveness of path guiding, we turn off the Next Event Estimation (NEE) for our and all comparison methods as done in previous work \cite{vorba2014line,muller2017practical}. Comparison results with NEE turned on are shown in the supplementary material.
The ground-truth images are rendered using path tracing with NEE for $2$ to $6$ days per scene.

\paragraph{Quantitative and qualitative evaluation.}
We now demonstrate the quantitative and qualitative results of our method and compare against other methods with equal rendering time.
For quantitative evaluation, we use the relative Mean Squared Error (rMSE, as used in \cite{Rath2020}) and the perceptually-based Structural Similarity Index (SSIM, as used in \cite{bako2019offline}) as metrics.  
Table.~\ref{tab:quant_rmse} shows the quantitative results of rMSEs and SSIMs of the full images of all 12 testing scenes.
The corresponding percentages of running time of sub-components are shown in Tab.~\ref{tab:runtime}. In most cases, the path and photon tracing on CPU take more than $90\%$ of the entire system running time, and we only spend a small amount of time ($10\%$) on requesting GPU resources for neural sampling map reconstruction. 
Our method achieves the best quantitative results with the lowest rMSEs and highest SSIMs on all testing images. 
Note that
ours is able to lower the rMSEs of the best comparison methods by more than $50\%$ in many challenging scenes like \ScnCE, \ScnVA, \ScnBR, \ScnH, \ScnSC, \ScnLR, and \ScnT.
These results demonstrate the high effectiveness and efficiency of our method, which is significantly better than all comparison methods.

To illustrate the details of our results, 
we also show quantitative and qualitative comparisons on multiple crops of the rendered images in Fig.~\ref{fig:teaser}, \ref{fig:transparent}, \ref{fig:complex_light} and \ref{fig:budget}.
Our results are of the highest visual quality in these figures, which can also be reflected by the lowest rMSEs of all the comparison image crops.

Note that the two unidirectional guiding methods \cite{muller2017practical, Rath2020} are usually the best two of all four comparison methods on these testing cases.
They utilize an adaptive tree as their spatial partitioning, which is more efficient than the uniform cache points used in \cite{vorba2014line},
leading to much better rendering quality in most testing scenes despite the fact that \cite{vorba2014line} is bidirectional.
However, it can be highly challenging for unidirectional methods to discover high-energy paths, when a scene involves complex specular-diffuse interactions (like those in Fig.~\ref{fig:transparent} that contain many reflective and refractive objects) or other strong global illumination effects (like in Fig.~\ref{fig:budget}).
Therefore, \cite{vorba2014line} sometimes achieves better results than the unidirectional ones, like the results of \ScnSS \ and \ScnLR, since it leverages photons from light paths that ease the process of light discovery.

In contrast, our approach also leverages an adaptive spatial structure and our novel hierarchical grid enables finer spatial partitioning than \cite{muller2017practical,Rath2020} in early iterations.
Meanwhile, our deep learning based method can reconstruct high-quality sampling maps from sparse photons; this enables high-quality path guiding in our finer spatial partitioning from the first through all iterations, avoiding the slow starting of those online learning methods and leading to highly efficient rendering.
Our approach purely relies on photons to reconstruct sampling maps, which is effective in general and also highly efficient for challenging scenes that are dominated by indirect lighting.
Thanks to our deep neural networks and our efficient spatial partitioning, 
our approach utilizes photons in a way that is much more efficient than previous work \cite{vorba2014line}. 
Our photon-driven neural path guiding approach enables high-quality rendering results that are significantly better than all previous unidirectional and bidirectional guiding methods.

\cite{bako2019offline} is a recent deep learning approach that first leverages an offline trained network for unidirectional path guiding; yet
their method can only guide the first bounces and leads to the worst results in most testing cases.
As shown in their paper, this technique can be effective for lowering the initial severe MC noise with sparse path samples, especially on scenes with strong direct illumination. 
However, such a first-bounce technique is not very effective for scenes with strong indirect illumination; the benefits of its offline learning also become more limited through longer rendering, once other traditional multi-bounce techniques get enough path samples online.
In contrast, our method is the first offline deep learning method that enables multi-bounce path guiding.
Our approach takes full advantage of an offline trained network and successfully models the incident light field at any local regions in a scene, enabling significantly better rendering quality than \cite{bako2019offline} and all other traditional multi-bounce guiding techniques.

\paragraph{Equal-quality comparison.}
In addition to the equal-time comparison, we also compare the time spent to achieve the results of similar quality on some highly challenging scenes shown in Fig.~\ref{fig:budget}; 
the corresponding rendering time (compared to our time) of each method is listed, for achieving the same rMSE (with a threshold of $10^{-4}$ in difference) of the full image as our method (corresponding to the rMSEs shown in Tab.~\ref{tab:quant_rmse}).
We can see that our method can significantly speed up the naive path tracing with the rendering speed that is several tens of times faster.
Moreover, the fastest comparison methods for these scenes still require at least two times the rendering time as our method does.
Our approach significantly reduces the required amount of time to achieve realistic rendering.

\paragraph{Sampling map reconstruction.}
The core of our path guiding approach is our deep learning based sampling map reconstruction.
We show examples of our reconstructed sampling maps, corresponding inputs and the ground-truth in Fig.~\ref{fig:map_vis}; more examples are provided in the supplementary material.
Note that our method can consistently improve the reconstruction quality through iterations. 
Even at the second iteration, when the input is extremely noisy, our network can still denoise the input and recover a full sampling map that has many details and is very close to the ground-truth.
We also show additional comparison with using a simple U-Net for sampling map reconstruction in the supplementary material.
This high-quality sampling map reconstruction allows for highly efficient path sampling when rendering.

To further justify the effectiveness of our network, we compare with only using the raw input sampling map (without the network reconstruction) for path guiding in Fig.~\ref{fig:ablation}. We also compare with a version that reconstructs sampling maps at a lower resolution of $64\times 32$ (we use $128\times 64$ by default as mentioned in Sec.~\ref{sec:impl}).
The results of \cite{muller2017practical} and \cite{Rath2020} (which generally performs the best among all comparison methods as stated) are also shown in the figure to better understand the position of these versions of our method with reduced or degraded components.
Note that, our method without the network can already achieve comparable rendering quality compared to previous methods in some cases. 
And for \ScnP, our method without network reconstruction can even perform better than \cite{muller2017practical}; this is because using photons is highly effective for such a scene, involving complex specular-diffuse interactions.
This example clearly demonstrates the benefit of leveraging photons.
The neural network in our framework can significantly improve the rendering quality achieved without the network. Our full model achieves the best visual quality and the lowest rMSE in these testing scenes.
Also note that, while worse than our final model, our method with a lower resolution of sampling maps can already outperform the comparison methods and the one without the network.
This demonstrates the high reconstruction quality of our network. This also shows that our method generalizes well on different sampling map resolutions, though a higher resolution often leads to higher quality.

\paragraph{Hierarchical grid.}
We now investigate our presented spatial structure - the hierarchical grid.
We show rMSEs of images rendered with different resolutions for the initial regular grid in Fig.~\ref{fig:grid}.
We also show corresponding results using only a regular grid without the adaptive partitioning inside voxels.
Note that, without the adaptive partitioning, rendering quality varies drastically across different resolutions, since a low-resolution grid lacks expressibility of complex light fields in the scene and a high-resolution grid does not have enough photons in each voxel.
On the contrary, our hierarchical grid is more stable with different resolutions, since it is able to adaptively subdivide the initial grid to a desired resolution locally.
Our hierarchical grid also consistently enables better rendering quality than a regular grid at the same initial resolution.

\paragraph{Temporal stability. }
We also evaluate the temporal stability of our method.
In particular, we use the DSSIM (i.e., dissimilarity as used in \cite{vogels2018denoising}) between consecutive frames with a moving camera to express the temporal stability. Figure~\ref{fig:temporal} shows the DSSIMs of our method and other comparison methods.
Since our renderings are consistently better than those of other methods, 
our method also achieves the best temporal stability.

\paragraph{Limitations.}
We use a regular grid to represent a sampling distribution as a standard 2D map (image). This is easy for a deep neural network to process and reconstruct. However, it consumes more memory than the directional quad-tree used in \cite{muller2017practical}; the memory also limits the resolution of sampling maps we can reconstruct. 
Nonetheless, we show that our resolution of $128\times 64$ can already achieve high-quality sampling, and even a lower resolution (like $64\times 32$ as shown in Fig.~\ref{fig:ablation}) can also provide better results than previous methods.
We leave extensions with a sparse directional representation in a learning framework as future work.
Our approach leverages photons to reconstruct sampling maps. However, tracing photons can sometimes be highly inefficient; for example, if a camera is looking at only a small region of a large scene, there can be a large number of photons that are traced but never reach any valid voxels, leading to very expensive photon tracing. Leveraging bidirectional guiding techniques like \cite{vorba2014line} to also guide the photon tracing process can potentially resolve this. 
Please refer to our supplementary material for an initial extension of our method with photon guiding. Finally, we currently use CPU for rendering and GPUs for neural reconstruction. Although we overlap data transfers with computation to reduce the synchronization latency, integrating our proposed framework to a GPU-based renderer like Nvidia OptiX \cite{parker2010optix} may be a better choice to accelerate the whole system.

\section{Conclusion and future work}
\label{sec:future}
In this paper, we present the first deep learning-based photon-driven path guiding approach.
Our approach leverages photons to reconstruct sampling distributions, which is more effective than pure unidirectional (path-driven) methods for challenging scenes that are dominated by indirect lighting; we propose to use a deep neural network to regress high-quality sampling maps from low-quality photon histograms, enabling highly efficient path guiding using only sparse photons.
To fully utilize the benefits of our network, we introduce an adaptive hierarchical grid to cache our reconstructed sampling maps across the scene, allowing for efficient path guiding at any spatial location.
We demonstrate that our method achieves significantly better quantitative and qualitative results than various previous state-of-the-art path guiding methods on diverse challenging scenes.

\begin{figure*}[t]
    \includegraphics[width=\linewidth]{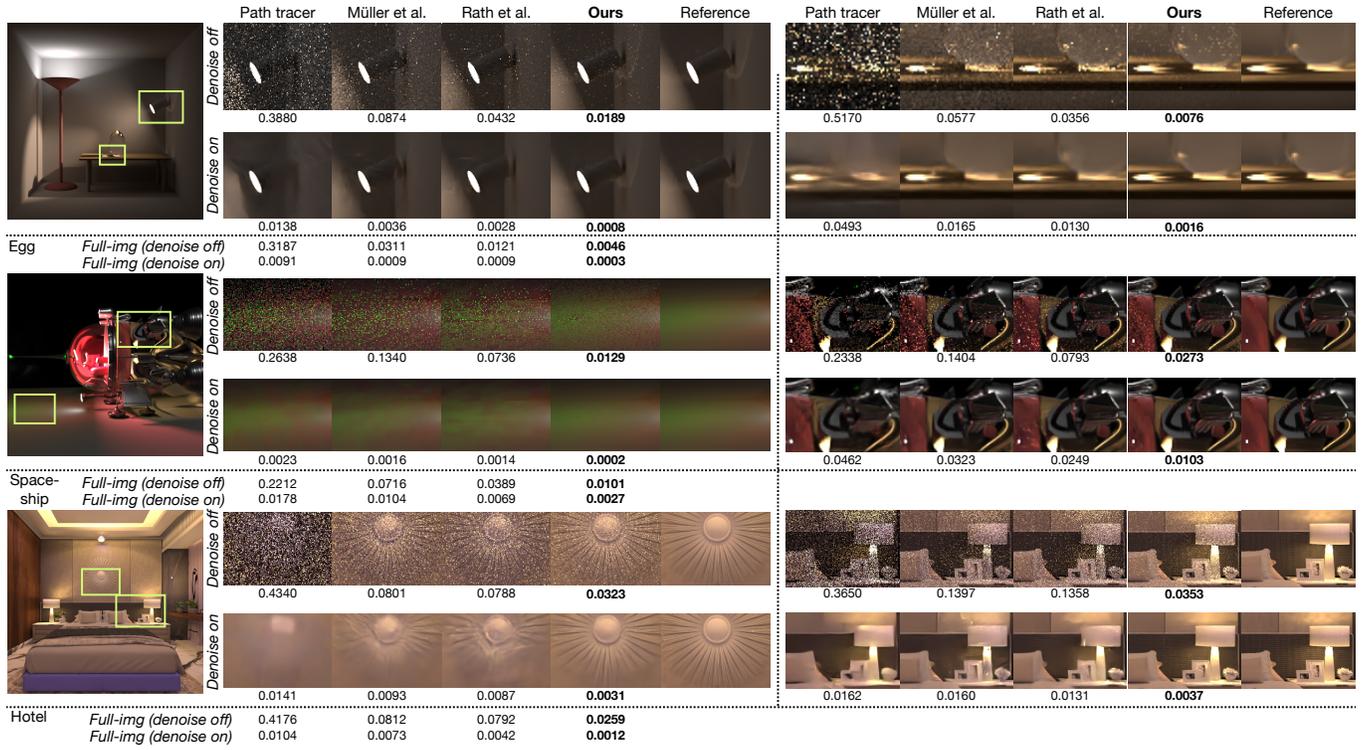}
    \caption{Monte-Carlo denoising on the path guiding rendering results. We use the default deep learning based denoiser in Nvidia OptiX 6.5. In general, the denoiser fills the holes in between pixels and filters out the high-frequency MC noise. The denoised images rendered with our method are more acceptable without severe blur or distortion.}
    \label{fig:denoise}
\end{figure*}

Our method is the first neural path guiding method that uses an offline trained network and supports guiding at any bounces, whereas previous related techniques either train an online network \cite{muller2019neural} or only guide the first bounce \cite{bako2019offline}.
We believe our method takes an important step towards making the neural path guiding more practical, thus also opening up many appealing future directions.
Our approach leverages local photon statistics for sampling map reconstruction; an interesting extension is to also consider some global context and even achieve global guiding in primary space (like \cite{muller2019neural,guo2018primary}).
In addition, our target sampling density function can be potentially extended to some advanced distributions like variance-aware \cite{Rath2020} or product sampling \cite{herholz2016product} (avoiding the one-sample MIS) for better sampling efficiency. 
Combining our deep learning based local sampling reconstruction with reinforcement learning techniques \cite{huo2020adaptive} to achieve sampling with a proper reward function could provide more benefits.
We leverage heuristic criteria to achieve voxel subdivision in the hierarchical grid; this spatial partitioning process could also be potentially learned via a deep neural network in the future.
While we purely leverage photons in our method, combining photons and path samples in a holistic neural path guiding framework is an interesting future direction to explore.
\Comment{\MM{I don't like the use of questions in the conclusion.  I'd try to be more declarative: "combining our deep learning reconstruction with reinforcement learning ... could provide more benefits"}}

\section*{Acknowledgements}
This work was supported in part by NSF grants 1703957 and 1764078, the Ronald L. Graham Chair, two Google Fellowships, an Adobe Fellowship and the UC San Diego Center for Visual Computing.

\section{Supplementary Material}
In this supplementary material, we provide additional experimental results and sampling map visualizations, as well as some discussions about potential extensions of our proposed framework. Although not being emphasized in the main paper, these additional studies and evaluations are also very important in the design of a full-fledged path guiding system in practice.

\subsection{Monte-Carlo Denoising}
Monte-Carlo (MC) rendering algorithms like path tracing are known to suffer from the slow convergence problem when producing noise-free images \cite{kajiya1986rendering, lafortune1996mathematical}. In recent years MC denoising has become a very successful approach to reduce pixel variance, especially those based on neural networks \cite{bako2017kernel, chaitanya2017interactive, vogels2018denoising}. Although MC denoising is a biased operation, it significantly increases the visual quality by removing the last-mile residual pixel noise. 
Therefore, we apply deep learning based denoising techniques on the rendered results from path guiding methods, which can be a practical way to use the method. In particular, we use the built-in denoiser from Nvidia OptiX 6.5 \cite{parker2010optix} to denoise the output of our method and baselines. Results in Fig.~\ref{fig:denoise} show that our method achieves the best performance even after denoising. In fact, although the denoiser can reduce rMSE almost in all cases, such denoising is only reasonable when the rendered image has a low level of noise; otherwise the results can appear blurry with missing details (e.g., the caustics part in the \textsc{Egg} scene) or distorted (e.g., the center of wall in the \textsc{Hotel} scene), which is not acceptable in either case for high-end production rendering.

\subsection{Next Event Estimation}
\begin{figure*}[t]
    \includegraphics[width=\linewidth]{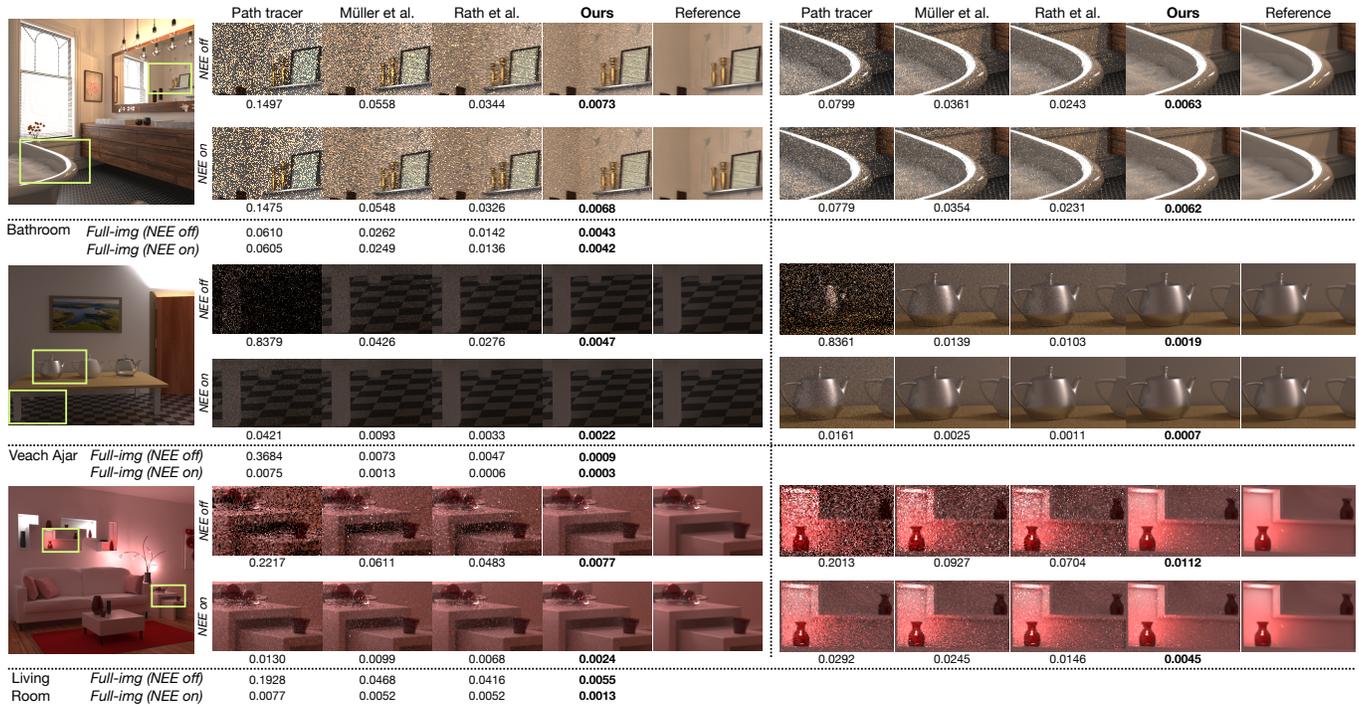}
    \caption{The effect of next event estimation (NEE) on the final rendering results. The comparison is equal-time for each row. When turning on the NEE, the rendering time increases since we keep a similar total sample count. Results show that NEE greatly improves the results in some cases while it is not very useful in some other cases, depending on the sampling map quality in path guiding as well as the levels of light visibility at different scene locations. }
    \label{fig:nee}
\end{figure*}

\begin{figure}[h]
    \includegraphics[width=\linewidth]{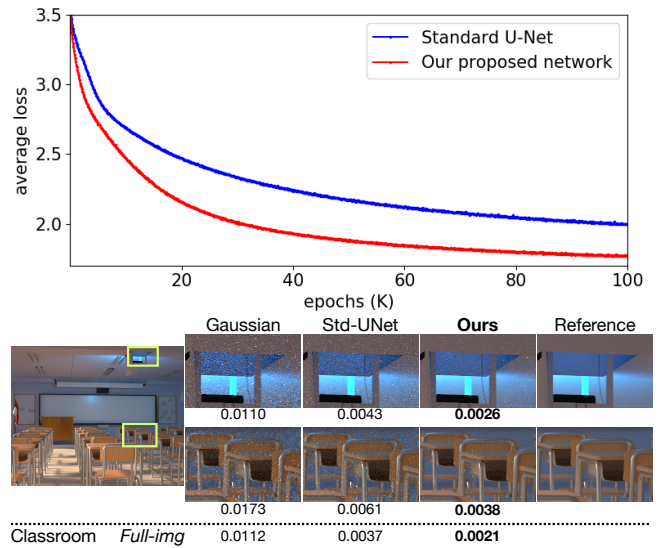}
    \caption{Our proposed neural network performs better than a single standard U-Net and traditional Gaussian filter in sampling map reconstruction, leading to lower-variance rendering results. The error curve is clipped for better visualization purposes.}
    \label{fig:supp_naive}
\end{figure}

In the default experimental setting, we turn off the next event estimation (NEE) to clearly compare the effects from path guiding (similar to \cite{muller2017practical} and \cite{vorba2014line}), though in practice NEE can be effective on some cases for all comparison methods. In particular, NEE can help reduce the variance by easing the search of a light source and improving the sampling map quality. 
To study how NEE affects the results, we turn on the NEE and keep the sample count unchanged on multiple test scenes. 
Results in Fig.~\ref{fig:nee} show that whether NEE is useful or not depends mostly on the light setup. 
For the \textsc{Bathroom} scene, the glass bulb fixture and staggered window blinds make the direct connection very hard to succeed; for the \textsc{Veach Ajar} and \textsc{Living Room} scenes, NEE fails and succeeds from time to time depending on the local light visibility. 
As a consequence, the rendering time greatly increases for all methods when NEE is turned on. 
In fact, our method achieves the best performance whether NEE is enabled or not, thanks to the high-quality reconstructed sampling maps that can capture both direct and indirect illumination. We believe the decision to request NEE or not is highly related to the total timing budget in specific applications.

\subsection{Neural Sampling Map Reconstruction}
\begin{figure*}[t]
    \includegraphics[width=0.95\linewidth]{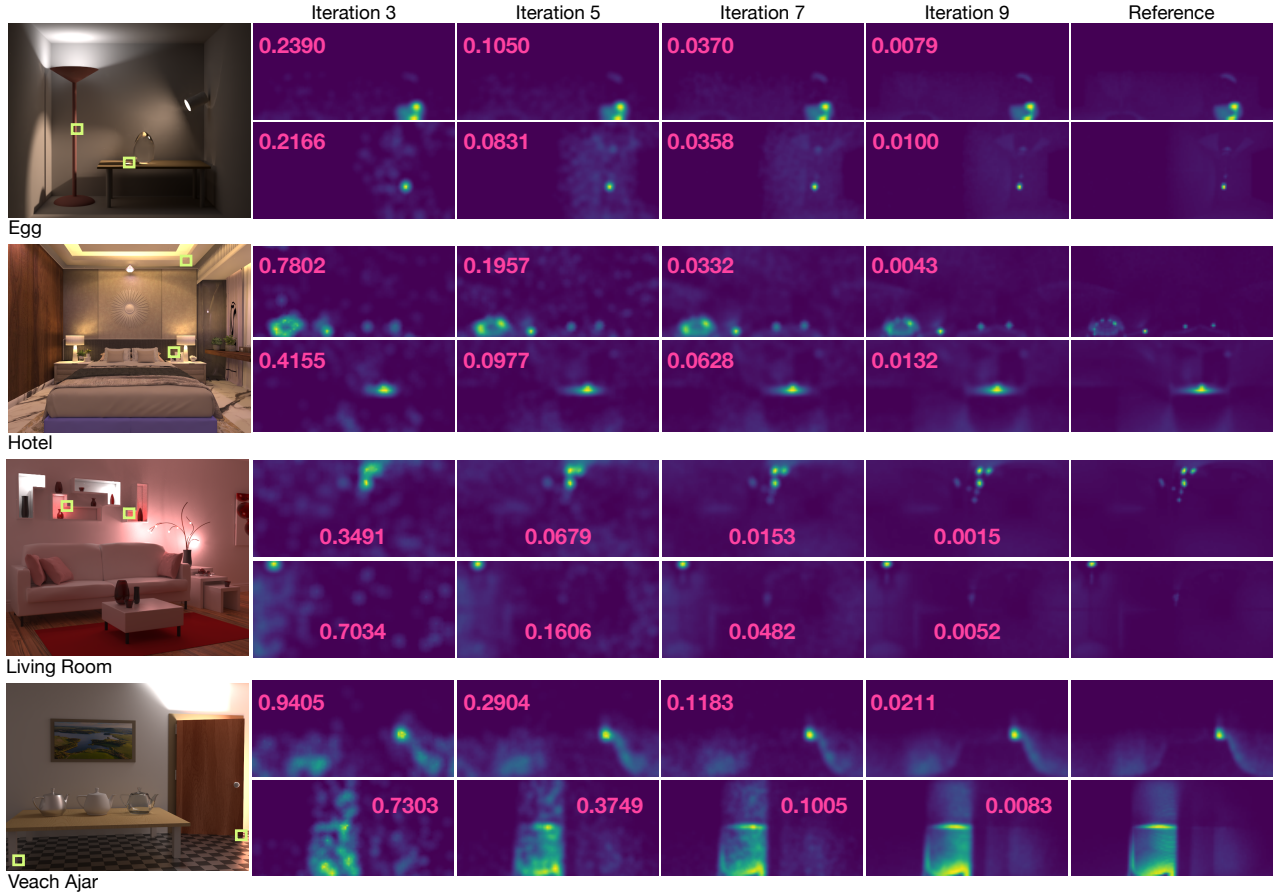}
    \caption{Additional reconstructed sampling map visualization through learning iterations. The reconstructed sampling maps lead to better path space exploration at the beginning and more accurate representations of the incident radiance in the subsequent iterations.}
    \label{fig:maps_extra}
\end{figure*}

\begin{figure}[h]
    \includegraphics[width=\linewidth]{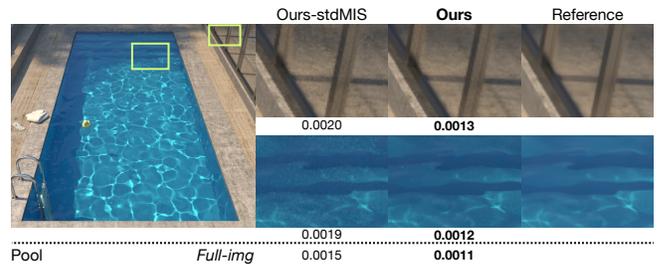}
    \caption{Our proposed heuristic one-sample MIS scheme performs better than the default mixture coefficient $\alpha=0.5$ especially when BSDF importance sampling and guiding have very disparate contributions to the final image.}
    \label{fig:supp_mis}
\end{figure}

As we mentioned in Sec.~6 of the main paper, we design a neural network that effectively reconstructs high-quality sampling maps. To demonstrate the effectiveness of our proposed network architecture, loss function, and our multi-expert inference scheme (Sec.~8 of the main paper), we train a single standard U-Net \cite{ronneberger2015u} with $l_{1}$ loss function and without the auxiliary features, and use this simplest network to reconstruct all the sampling maps without training multiple versions. In addition, we try a simple Gaussian filter denoising and choose the best result from a range of variances from 0.01 to 10. Figure.~\ref{fig:supp_naive} shows the error curve during neural network training, as well as a visual comparison on the \textsc{Classroom} scene. Although deep learning based results are both better than the one without applying neural reconstruction, our proposed neural networks can produce a more smooth and lower-noise image. As shown in the loss curve, the average error of our reconstructed sampling maps is also smaller. The traditional Gaussian filter gives much worse performance since it only adds the same level of blur to the entire sampling map. We believe our proposed neural network can be further compressed by the state-of-the-art network compression methods \cite{cheng2018model, deng2020model} and improved by more advanced architectures in the future.

\subsection{One-Sample MIS}
In Sec.~7.3 of the main paper, we demonstrate a new heuristic pipeline for estimating mixture coefficient $\alpha$ in one-sample MIS of BSDF sampling and guiding. This is quite useful in some cases, as shown in Fig.~\ref{fig:supp_mis}. For example in the \textsc{Pool} scene, the BSDF sampled directions from the floor often fail to find the light source and leave the scene permanently, leading to a small contribution to the final pixel color. In contrast, our heuristic encourages sending more guiding samples in those regions based on the statistics of previously traced path samples. And for very glossy surfaces such as the metal armrest in this scene, we send more BSDF samples since many guided directions can have very small or zero BSDF value. Although the proposed heuristic may be sub-optimal, it is straightforward to implement and does not introduce extra online optimization overhead. In the future, we believe our heuristic can provide a good starting point to initialize other methods that try to optimize $\alpha$ in path guiding \cite{mueller19guiding, Rath2020}. 

\begin{figure}[t]
    \includegraphics[width=\linewidth]{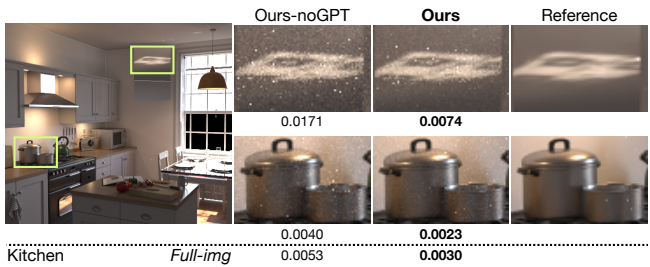}
    \caption{We study the effectiveness of guided photon tracing extension (GPT). The reconstructed sampling maps have lower quality when photons are too sparse since there is not enough information for rebuilding the incident radiance distribution. In contrast, it is sometimes beneficial to guide traced photons into visually important regions. }
    \label{fig:gpt}
\end{figure}

\subsection{Guided Photon Tracing}
We leverage photons in our neural path guiding, which is very effective for dominant indirect lighting. 
However, photons are only useful when they are visible to the camera and in some cases many wasted photons can be traced. 
In order to handle some special light transport cases where many photons are invisible, we add the guided photon tracing module to our system as a simple extension. 
Similarly to \cite{vorba2014line}, we reconstruct the importance sampling maps from the accumulated path samples in each voxel. These path samples are virtual particles (i.e., importons \cite{peter1998importance}) containing a value describing with what factor an illumination at a certain location would contribute to the final image. 
Here, we use the same pre-trained network for such reconstruction. 
The reconstructed importance sampling maps are then used for guiding photon tracing in every learning iteration. We use the \textsc{Kitchen} scene as an example since many emitted photons from outside sunlight cannot land inside the room without guided photon tracing, unless they are explicitly programmed to do so by manually providing the location of windows as our default experimental setting. Figure.\ref{fig:gpt} shows the reduced variance in regions that are indirectly illuminated by the white sunlight and comparable results on regions that are lit by the indoor orange-color lights. Apart from these special cases, it is not always necessary to add this extra module for most common lighting setups created by the lighting artists when most photons are visible to the camera. Besides, our neural network is trained to properly handle the input maps with multiple levels of sparsity so our system can work well as long as the photons are not too sparse.

\subsection{Additional Sampling Map Visualization}
Some additional sampling maps are visualized in Fig.~\ref{fig:maps_extra}. After pre-training on an offline dataset, our neural network can progressively reconstruct higher-quality sampling maps with more accumulated photon energies through iterations on new scenes. Unlike previous Monte-Carlo denoising networks \cite{chaitanya2017interactive, bako2017kernel, vogels2018denoising} which only process the input image once and stop after the inference, our reconstruction is getting better and closer to the ground-truth sampling maps over time. More specifically, the network reconstructs blurrier sampling maps due to low confidence in the early iterations to encourage more exploration of the directional space for the following path tracing; in the later iterations the reconstructed sampling maps get sharper and there emerge more accurate details of the incident radiance distribution due to a higher level of confidence.
\bibliographystyle{ACM-Reference-Format}
\bibliography{ref}


\end{document}